\documentclass[11pt]{article}
\usepackage{epsf}
\usepackage{color}
\begin{document}

\def\slash#1{{\rlap{$#1$} \thinspace/}}

\begin{flushright} 
{\sf February, 2004 } \\
{\sf KEK-TH-938}
\end{flushright} 

\vspace{0.1cm}

\begin{Large}
\vspace{1cm}
\begin{center}
{{\sf Nonabelian gauge field 
and dual description of fuzzy sphere }
}   \\
\end{center}
\end{Large}

\vspace{1cm}

\begin{center}
{\large {\sf Yusuke Kimura} }  \\ 

\vspace{0.5cm} 
{\sf Theory Group, KEK, }\\
{\sf Tsukuba, Ibaraki 305-0801, Japan} \\

\vspace{0.5cm} 
{\sf kimuray@post.kek.jp}
\vspace{0.8cm}

\end{center}

\vspace{1cm}

\begin{abstract}
\noindent
\end{abstract}
\hspace{0.4cm}
{\sf

In matrix models, higher dimensional D-branes are 
obtained by imposing a noncommutative relation 
to coordinates of lower dimensional D-branes. 
On the other hand, a dual description 
of this noncommutative space 
is provided by higher dimensional 
D-branes with gauge fields. 
Fuzzy spheres can appear as 
a configuration of lower dimensional D-branes 
in a constant R-R field strength background. 
In this paper, we consider a dual description of 
higher dimensional fuzzy spheres 
by introducing nonabelian gauge fields 
on higher dimensional spherical D-branes. 
By using the Born-Infeld action, 
we show that a fuzzy $2k$-sphere and 
spherical D$2k$-branes with a nonabelian gauge field 
whose Chern character is nontrivial 
are the same objects 
when $n$ is large. 
We discuss a relationship between the 
noncommutative geometry and nonabelian gauge fields. 
Nonabelian gauge fields are represented by noncommutative 
matrices including the coordinate dependence. 
A similarity to the quantum Hall system is also studied. 
}

\newpage 


\vspace{1cm}

\section{Introduction}
\hspace{0.4cm}
String theory is considered as the best candidate for 
a quantum theory of gravity or 
a unified theory of all interactions. 
Much attention has been given to noncommutative geometry 
because it is expected to capture 
some aspects of the quantum gravity. 
Over the past few years 
several papers have been devoted to the study of 
a relationship between noncommutative geometry and string theory. 
The need of noncommutative geometry in string theory is easily 
understood by considering a world-volume action of D-branes. 
D-branes are defined as the endpoints of open strings. 
Since gauge fields appear in the ground state of 
open strings, the low energy dynamics of D-branes is described 
by gauge fields. 
One of the most interesting aspects is the appearance 
of nonabelian gauge symmetry from the world-volume 
theory of some coincident D-branes. 
A low energy effective action of $N$ D-branes is provided 
by the dimensional reduction of the ten-dimensional 
$U(N)$ Yang-Mills theory, 
and transverse coordinates of $N$ D-branes are expressed 
by $U(N)$ adjoint scalars \cite{witten}. 
The appearance of this matrix-valued coordinate 
indicates relationships between string theory and 
noncommutative geometry. 

Concrete models for studying the nonperturbative 
properties of string theory are proposed in \cite{BFSS,IKKT}. 
These matrix models are constructed 
by taking lower dimensional D-branes as 
fundamental degrees of freedom. 
Since higher dimensional D-branes also exist in string 
theories, they need to be contained in matrix models. 
It is important to understand relationships between D-branes of 
different dimensions because higher dimensional ones 
are constructed form lower dimensional ones. 
A study of a world-volume action for D-branes 
can reveal the relationship.  
The Chern-Simons coupling indicates an 
interesting relationships between D-branes 
\cite{hep-th/9510161,hep-th/9512077,hep-th/9910052,myers}. 
Let us consider a matrix model of D$0$-branes 
as a simple example. 
In this model, a D$2$-brane 
is represented by two noncommutative matrices, 
\begin{equation}
[X_{1},X_{2}]=-iC_{12}1,  
\label{noncommutativeplane} 
\end{equation}
where $X_{i}$ are transverse coordinates of D$0$-branes 
\cite{townsend,BFSS}. 
The commutator measures the charge of D$2$-brane. 
In general, we can say that 
higher dimensional D-branes are expressed by noncommutative 
geometry from lower dimensional D-branes. 
What has to be noticed is 
that the above matrix configuration is not just a D$2$-brane but a bound 
state of D$0$-branes and a D$2$-brane. 
In the world-volume theory of a D$2$-brane, 
D$0$-branes bounded on it 
are expressed by an abelian gauge field configuration 
with a nonzero first Chern class, which is supported 
by the following coupling: 
\begin{equation}
\mu_{2}\int_{2+1} \left(C^{(3)}+ C^{(1)}\wedge F\right), 
\end{equation}
where $C^{(k)}$ represents R-R $k$-form field. 
This coupling implies that a gauge field on 
a D$2$-brane couples to the R-R one-form field, 
which is associated with D$0$-branes. 
Therefore the first Chern class provides the charge of D$0$-branes. 
This fact can be generalized to the fact that 
lower dimensional D-branes bounded on higher dimensional 
D-branes are represented by gauge fields 
which is topologically nontrivial. 
We then find that there are two descriptions for a system. 
If we start with lower dimensional D-branes, higher dimensional 
D-branes are obtained by imposing a noncommutative relation 
to coordinates of lower dimensional D-branes. 
On the other hand, 
on the world-volume theory of higher dimensional D-branes, 
a gauge field with a nonvanishing Chern number 
is needed to express lower dimensional D-branes. 
See also \cite{ishibashi} for this correspondence. 
These facts are extensively reviewed in 
\cite{polreview,taylorreview,hep-th/0303072}. 

In the previous paragraph, we showed an example of two 
dual descriptions for a bound state of 
a flat D$2$-brane and D$0$-branes. 
A noncommutative plane is dual to a flat D$2$-brane 
with an abelian gauge field. 
It is important to study two descriptions 
because the second description 
gives another viewpoint for noncommutative geometry. 
This study helps us to understand 
the role of noncommutative geometry in string theory.  
This correspondence is also expected to hold for curved D-branes. 
A simple generalization is to consider the fuzzy sphere. 
A fuzzy two-sphere has a dual description in terms of 
an abelian gauge field on a spherical D$2$-brane, 
and is interpreted as a bound state of a spherical D$2$-brane 
and D$0$-branes 
\cite{kabattaylor,myers,hep-th/9911136,hep-th/0101211,
hep-th/0212183,hep-th/0302190}. 
We need to emphasize that 
gauge fields are abelian for both cases. 
A higher dimensional generalization is an interesting 
subject since a higher dimensional fuzzy sphere 
has some characteristic features. 
It is first constructed in \cite{hep-th/9712105}, 
and further analyses revealed the interesting structures 
of higher dimensional fuzzy sphere 
\cite{hep-th/0105006,hep-th/0111278,hep-th/0301055}. 
See also \cite{hep-th/0204256,azumabagnoud,
hep-th/0210166,hep-th/0212214,hep-th/0312190}. 
It is pointed out in \cite{hep-th/0111278} that 
a fuzzy $2k$-sphere is expressed by a coset space 
$SO(2k+1)/U(k)$, and therefore the dimension is 
not $2k$ but $k(k+1)$. 
The number of the extra dimensions makes the higher dimensional 
fuzzy sphere complicated. 
As analyzed in \cite{hep-th/0102080} for the $k=2$ case, 
a nonabelian gauge field is needed to realize 
a dual description of higher dimensional fuzzy spheres. 
The use of a nonabelian gauge field is closely related 
to the existence of the extra dimensions. 
A dual aspect of higher dimensional fuzzy spheres 
is partially discussed in 
\cite{hep-th/0102080,hep-th/0111278,hep-th/0201016,
hep-th/0306250}.  
In this paper, we explicitly construct a 
nonabelian gauge field to realize a dual description 
of higher dimensional fuzzy spheres 
and compare two descriptions by using the Born-Infeld 
action. We see that two descriptions provide 
the same result when the size $n$ of matrices 
realizing the fuzzy sphere is enough large. 
Such a analysis shows an interesting relationship between 
the noncommutativity on the fuzzy sphere and 
the nonabelian gauge field. 
See also a recent work \cite{hep-th/0401043} 
which calculates D-brane charges for various 
noncommutative configurations. 

The organization of this paper is as follow. 
In section \ref{sec:fuzzysphere}, 
we review the algebra of the fuzzy $2k$-sphere. 
Some distinctive aspects of the higher dimensional 
fuzzy sphere are explained. 
We consider $N$ D$0$-branes in a constant R-R $(2k+2)$-form 
field strength background in section \ref{sec:DbraneunderRR}. 
An action of D$0$-branes in the background is described by 
a matrix model with the Chern-Simons term. 
We see that $N$ D$0$-branes form a fuzzy $2k$-sphere at 
a classical extremum of the matrix model. 
This phenomenon is called the dielectric effect \cite{myers}. 
We calculate the value of the potential for the fuzzy sphere. 
Since higher dimensional fuzzy spheres are realized 
at a local maximum of the classical potential, 
they are classically unstable in this situation. 
To understand a dual description of higher dimensional fuzzy spheres 
is a main part of this paper. 
We consider a world-volume theory of spherical D$2k$-branes 
in section \ref{sec:dualdescription}. 
A dual description of a fuzzy $2k$-sphere is expected to be given by 
a bound state of spherical D$2k$-branes and D$0$-branes. 
The D$0$-branes are represented by a gauge field configuration 
with a nonzero Chern number. 
Such gauge fields are recently used to construct 
a higher dimensional quantum Hall system 
in \cite{zhanghu,cond-mat/0306045,cond-mat/0306351,hep-th/0310274}. 
An interesting fact for the higher dimensional case 
is that the gauge field is nonabelian and 
expressed by using matrices which are associated with 
a lower dimensional fuzzy sphere. 
This was also suggested in \cite{hep-th/0301055} 
by studying a relationship 
between a higher dimensional fuzzy sphere 
and a lower dimensional fuzzy sphere. 
A world-volume action of spherical D$2k$-branes 
with a (nonabelian) gauge field shows two extrema, 
one of them corresponds to a fuzzy $2k$-sphere in the matrix 
model description. 
We compare some quantities such as the potential, 
the radius of sphere and the charge of lower dimensional 
D-branes in two descriptions, 
and it is shown that 
these two descriptions coincide at large $n$. 
Accordingly we conclude that 
a fuzzy $2k$-sphere and spherical D$2k$-branes 
with an $SO(2k)$ nonabelian gauge field 
are the same objects at large $n$. 
In section \ref{sec:noncommutativegaugefield}, 
we discuss a relationship between the noncommutativity and 
nonabelian gauge fields. 
It is explained that a nonabelian gauge field 
is described by a matrix including the coordinate 
dependence. 
We can observe an interesting mixing between 
the noncommutativity of coordinates and that of 
nonabelian gauge fields. 
We also provide an explanation about 
the large $n$ limit by studying the lowest Landau level condition 
in a higher dimensional quantum Hall system. 
A relation to the zero slope limit in \cite{SW} 
is also discussed. 
Section \ref{sec:summarydiscussions} devotes 
to summary and discussions. 
In appendix \ref{sec:someformulaefuzzysphere}, 
we summarize some formulae of the fuzzy sphere. 
In appendix \ref{sec:hopfmapandberryphase}, 
we review the construction of monopole 
gauge fields on even dimensional spheres. 

\section{Fuzzy Sphere}
\label{sec:fuzzysphere}
\hspace{0.4cm}
In this section, we review the algebra of fuzzy sphere in diverse 
dimensions. 
Our interest in this paper is restricted to 
even dimensional spheres.  
Odd-dimensional fuzzy spheres are investigated in 
\cite{hep-th/0101001,hep-th/0207111}. 

\vspace{0.4cm}

We first explain the fuzzy two-sphere 
\cite{hoppethesis,madore,kabattaylor,watamura,IKTW}. 
A coordinate of a fuzzy two-sphere is given by the $SU(2)$ algebra;
\begin{equation}
[X_{\mu},X_{\nu}]=2i\alpha
\epsilon_{\mu\nu\lambda}X_{\lambda}, 
\hspace{0.4cm}X_{\mu}=\alpha L_{\mu}, 
\label{algebraoftwosphere}
\end{equation}
where $L_{\mu}$ denotes the spin $n/2$-representation of 
$SU(2)$, and $\alpha$ is a dimensionful constant. 
$n$ can take any positive integers. 
The quadratic Casimir of $SU(2)$ provides the radius of 
the two-sphere;
\begin{equation}
r^{2}\equiv X_{\mu}X_{\mu}=
\alpha^{2}L_{\mu}L_{\mu}
=\alpha^{2}n(n+2)1_{n+1}.  
\end{equation}
In this realization, the size $(n+1)$ of the matrix is interpreted 
as the number of D$0$-branes. 

Let us now consider the place which is labelled by 
$L_{3}=n$, which corresponds to the north pole of 
a two-sphere. Around this point, the fuzzy two-sphere algebra 
(\ref{algebraoftwosphere}) becomes a noncommutative plane 
after we take a large $n$ limit, 
\begin{equation}
[X_{\mu},X_{\nu}]=2i\alpha^{2}  n\epsilon_{\mu\nu}1. 
\end{equation}
In this sense, the fuzzy sphere algebra is considered as 
a generalization of a noncommutative plane.

\vspace{0.4cm}

We next review the algebra of higher dimensional fuzzy spheres 
\cite{hep-th/9712105,hep-th/0105006,hep-th/0111278,hep-th/0301055}. 
It is natural to start with the $SO(2k+1)$ algebra since 
it is a symmetry of a $2k$-sphere, 
\begin{equation}
[\hat{G}_{\mu\nu},\hat{G}_{\lambda\rho}]=2\left(
\delta_{\nu\lambda}\hat{G}_{\mu\rho}
+\delta_{\mu\rho}\hat{G}_{\nu\lambda}
-\delta_{\mu\lambda}\hat{G}_{\nu\rho}
-\delta_{\nu\rho}\hat{G}_{\mu\lambda}
\right). 
\label{so(2k+1)algebra}
\end{equation}
An important fact is that 
any representations of this algebra do not 
always construct a fuzzy sphere. 
We have to consider a representation 
whose highest weight state is labelled by 
$[0,\cdots,0,n]$, where $n$ is a positive integer. 
We have used the Dynkin index to label the representation. 
This representation is considered as 
a generalization of the spinor representation since 
$\hat{G}_{\mu\nu}$ reduces to $\Gamma_{\mu\nu}$ when $n=1$. 
Note that $\hat{G}_{\mu\nu}$ is explicitly constructed 
from $\Gamma_{\mu\nu}$ 
by using a symmetric tensor product as in 
\cite{hep-th/9712105,hep-th/0105006} 
(see also \cite{hep-th/0401120} for the detailed calculation). 

We denote the size of the matrix representation by $N_{k}$. 
It is calculated as \cite{hep-th/0105006}:
\begin{eqnarray}
&&N_{1}=n+1, \hspace{0.4cm}
N_{2}=\frac{1}{6}(n+1)(n+2)(n+3), \cr 
&&N_{3}=\frac{1}{360}(n+1)(n+2)(n+3)^{2}(n+4)(n+5), \cr
&&N_{4}=\frac{1}{302400}(n+1)(n+2)(n+3)^{2}
(n+4)^{2}(n+5)^{2}(n+6)(n+7). 
\label{sizefuzzysphere}
\end{eqnarray}
The $k=1$ case is included in this table. 
A big difference between a fuzzy two-sphere and 
a fuzzy $2k$-sphere ($k\neq1$) is that 
$N_{1}$ can take any positive integers while $N_{k}$ $(k\neq1)$ cannot. 
We provide some comments on this point 
in section \ref{sec:summarydiscussions}.

Coordinates of a $2k$-sphere are included in $\hat{G}_{\mu\nu}$
\footnote{$i\hat{G}_{5a}$ are naturally identified with coordinates 
around the north pole. }. 
It is convenient to introduce $\hat{G}_{\mu}$ which satisfy 
\begin{eqnarray}
&&
[\hat{G}_{\mu},\hat{G}_{\nu}] =
2\hat{G}_{\mu\nu}, \cr
&&[\hat{G}_{\mu},\hat{G}_{\nu\lambda}]=2\left(
\delta_{\mu\nu}\hat{G}_{\lambda}-\delta_{\mu\lambda}
\hat{G}_{\nu}
\right).
\label{fuzzyspherealgebra2}
\end{eqnarray}
When $n=1$, $\hat{G}_{\mu}$ becomes the $(2k+1)$-dimensional 
gamma matrix $\Gamma_{\mu}$. 
We can define coordinates of a fuzzy $2k$-sphere as 
$X_{\mu}=\alpha\hat{G}_{\mu}$ 
because of the following relation 
\begin{equation}
\hat{G}_{\mu}\hat{G}_{\mu}=n(n+2k) 1_{N_{k}}. 
\label{radiusofsphere}
\end{equation}
The radius of a fuzzy $2k$-sphere is 
\begin{equation}
r^{2}=\alpha^{2}n(n+2k). 
\label{radiusof2ksphere}
\end{equation}
Other relations for $\hat{G}_{\mu}$ and $\hat{G}_{\mu\nu}$ 
are summarized in appendix \ref{sec:someformulaefuzzysphere}.

$\hat{G}_{\mu}$ and $\hat{G}_{\mu\nu}$ really form 
the $SO(2k+2)$ algebra. 
If we define $\Sigma_{\mu\nu}\equiv\hat{G}_{\mu\nu}$ and 
$\Sigma_{2k+2,\mu}\equiv i\hat{G}_{\mu}$, $\Sigma_{ab}$ 
$(a,b=1,\cdots,2k+2)$ satisfy the $SO(2k+2)$ algebra, 
belonging to an irreducible spinor representation of 
$SO(2k+2)$.  

Let us define a noncommutative scale on a fuzzy sphere. 
A commutator of coordinates is given by 
$\alpha^{2}\hat{G}_{\mu\nu}$, and the order of $\hat{G}_{\mu\nu}$ 
is $n$ due to the relation (\ref{GmunuGmunu=nn}). 
Therefore a noncommutative scale $l_{nc}$ can be defined 
as follows, 
\begin{equation}
l_{nc}^{2} \simeq \alpha^{2}n \simeq r^{2}/n. 
\label{noncommutativescale}
\end{equation}

We now comment on the number of independent matrices. 
$\hat{G}_{\mu}$ and $\hat{G}_{\mu\nu}$ have 
$(2k+1)$ and $k(2k+1)$ components respectively. 
Although there are totally $(k+1)(2k+1)$ components, 
each component is not independent because of some constraints 
between them. The number of independent 
components is really given by $k(k+1)$. This is understood 
by considering the fact that the fuzzy $2k$-sphere is 
given by the coset space $SO(2k+1)/U(k)$ \cite{hep-th/0111278}. 
The difference between 
the fuzzy sphere which is given by $SO(2k+1)/U(k)$ and 
the usual sphere by $SO(2k+1)/SO(2k)$ 
is important. 
It is shown in \cite{hep-th/0111278} that 
a fuzzy sphere has a bundle structure over a usual sphere. 
Accordingly higher dimensional fuzzy spheres 
have some extra dimensions. 
Coordinates of the extra dimensions are also noncommutative, 
and a noncommutative scale of the extra space 
is also given by (\ref{noncommutativescale}). 
In the remainder of this paper, we clarify this unusual aspect 
by considering the dual description of the fuzzy sphere.


\section{D$0$-branes under R-R field strength background}
\label{sec:DbraneunderRR}
\hspace{0.4cm}
In this section, we consider a collection of D$0$-branes 
in a constant R-R field strength background. 
A coupling of D$0$-branes with the R-R field strength background 
induces an interesting effect. 
D$0$-branes are polarized into a higher dimensional 
noncommutative geometry, which is 
called the dielectric effect \cite{myers}. 
A low energy dynamics of D$0$-branes in such a background 
is described by a matrix action. We will see that 
transverse coordinates of D$0$-branes form a fuzzy 
sphere at an extremum of the model. 
Some works investigating higher dimensional 
D-branes in such a background 
have been reported in \cite{hep-th/0007011,hep-th/0106253}. 

The tension and the charge of a D$p$-brane are defined as 
\begin{equation}
T_{p}=\mu_{p}=\frac{2\pi}{g(2\pi l_{s})^{p+1}}. 
\end{equation} 
We also define $\lambda \equiv 2\pi l_{s}^{2}$. 
In this notation, the charge of a D$p$-brane is related to 
that of a D$0$-brane as 
$\mu_{0}=(2\pi\lambda)^{p/2}\mu_{p}$. 
Basically we follow the notation of \cite{myers}. 

\vspace{0.2cm}

The dynamics of a world-volume theory of D-branes 
is described by the Born-Infeld action \cite{leigh} 
and the Chern-Simons action \cite{hep-th/9510161,hep-th/9512077}. 
The case of nonabelian was developped in \cite{hep-th/9910052,myers}. 
The Born-Infeld action for $N$ D$0$-branes 
in a flat space, with all other fields 
except transverse scalar fields 
vanishing, 
is given by 
\begin{eqnarray}
S_{BI}=-T_{0}\int dt Tr\sqrt{det(Q^{i}_{j})}
\simeq -T_{0}\int dt \left(
N-\frac{\lambda^{2}}{4}Tr[\Phi_{i},\Phi_{j}][\Phi_{i},\Phi_{j}]
\right), 
\label{BIpart}
\end{eqnarray}
where we have expanded the square root by assuming the condition 
$\lambda [\Phi_{i},\Phi_{j}] \ll 1$. 
This action can describe 
a low energy dynamics of $N$ D$0$-branes in a flat space. 
$\Phi_{i}$ is an $N\times N$ matrix-valued coordinate 
whose dimension is $length^{-1}$, 
representing a transverse motion of $N$ D$0$-branes. 
We define 
a coordinate whose dimension is $length$ as 
$X_{i}=\lambda \Phi_{i}$. 
The second term in (\ref{BIpart}) 
is also obtained form the dimensional 
reduction of the ten-dimensional $U(N)$ super Yang-Mills action. 

We next consider the Chern-Simons term. 
A coupling of $N$ D$0$-branes to the R-R potential 
is given \cite{hep-th/9910052,myers} by
\begin{eqnarray}
S_{CS}&=& \mu_{0}\int Tr 
\left(P[e^{i\lambda i_{\Phi} i_{\Phi}} \sum C^{(n)}] \right)\cr 
&=&
\mu_{0}\int Tr \left(
P\left[C^{(1)} +i\lambda i_{\Phi}i_{\Phi}C^{(3)}
-\frac{\lambda^{2}}{2}(i_{\Phi}i_{\Phi})^{2}C^{(5)} 
\right.\right. \cr
&& \hspace{3cm} \left.\left. 
-i\frac{\lambda^{3}}{3!}(i_{\Phi}i_{\Phi})^{3}C^{(7)}
+\frac{\lambda^{4}}{4 !}(i_{\Phi}i_{\Phi})^{4}C^{(9)}
\right]\right). 
\label{CSterm}
\end{eqnarray}
In this action, we considered a case where 
the NS-NS two-form field $B$ and 
the gauge field strength $F$ vanish. 
The notation $P[\cdots]$ denotes the pull back of 
the spacetime tensor to the world-volume, 
and acts on $C^{(1)}$ as
\begin{equation}
P[C^{(1)}]_{t}=C_{\mu}
\frac{\partial X_{\mu}}{\partial t}
=C_{t}+\lambda C_{i}
\frac{\partial \Phi_{i}}{\partial t}. 
\end{equation}
We must draw attention to the fact that 
the R-R fields are functional of 
transverse scalar fields $\Phi$. 
The dependence is actually defined by using a non-abelian 
Taylor expansion \cite{hep-th/9809100}; 
\begin{equation}
C(t,\Phi)=
\sum_{n=0}^{\infty}\frac{\lambda^{n}}{n!}
\Phi^{i_{1}}\cdots \Phi^{i_{n}}
(\partial_{x_{i_{1}}}
\cdots \partial_{x_{i_{n}}})
C(t,x)\mid_{x=0}. 
\end{equation}
The symbol $i_{\Phi}$ is introduced in \cite{myers}, 
and denotes the following operation, 
\begin{equation}
i_{\Phi}i_{\Phi}C^{(2)}=\Phi^{j}\Phi^{i}C_{ij}^{(2)}. 
\end{equation}
We now consider a case where only 
a constant R-R $(2k+2)$-form field strength 
is nonzero: 
\begin{equation}
F^{(2k+2)}_{ti_{1}\ldots i_{2k+1}}=
f_{k} \epsilon_{i_{1}\ldots i_{2k+1}}, 
\label{RR2k+2}
\end{equation}
where $f_{k}$ is a constant and determined later. 
Then we find that the leading order interaction term is obtained 
from the Chern-Simons term (\ref{CSterm}) as 
\begin{eqnarray}
S_{CS}&=&-\frac{1}{(2k+1)}\frac{i^{k}}{k!}\lambda^{k+1}\mu_{0}
\int dt Tr(\Phi_{i_{1}}\cdots\Phi_{i_{2k+1}})
F^{(2k+2)}_{ti_{1}\ldots i_{2k+1}} \cr 
&=&-\frac{1}{(2k+1)}\frac{i^{k}}{k!}\lambda^{k+1}f_{k}\mu_{0}
\int dt Tr(\Phi_{i_{1}}\cdots\Phi_{i_{2k+1}})
\epsilon_{i_{1}\ldots i_{2k+1}}. 
\label{CSpart}
\end{eqnarray}

By combining (\ref{BIpart}) with (\ref{CSpart}), 
a low energy effective scalar potential of $N$ D$0$-branes 
in a constant R-R $(2k+2)$-form field strength background 
is found to be 
\begin{equation}
V=\lambda^{2}T_{0}\left(
-\frac{1}{4}Tr[\Phi_{i},\Phi_{j}][\Phi_{i},\Phi_{j}]
+\frac{1}{(2k+1)}\frac{i^{k}}{k!}\lambda^{k-1}f_{k}
 Tr(\Phi_{i_{1}}\cdots\Phi_{i_{2k+1}})
\epsilon_{i_{1}\ldots i_{2k+1}}
\right), 
\label{D0matrixmodelinRR}
\end{equation}
where we have ignored the energy of $N$ D$0$-branes, $NT_{0}$, 
which is not important in this discussion. 
The indices $i,j$ run over $1,2,\cdots,2k+1$. 
$f_{k}$ is determined by requiring the condition that 
a fuzzy $2k$-sphere becomes a classical solution of 
this matrix model. 
The equation of motion for 
the matrix model (\ref{D0matrixmodelinRR}) is 
\begin{equation}
[\Phi_{j},[\Phi_{i},\Phi_{j}]]
-\frac{i^{k}}{k!}\lambda^{k-1}f_{k}
\Phi_{i_{2}}\cdots\Phi_{i_{2k+1}}
\epsilon_{ii_{2}\ldots i_{2k+1}}
=0.
\end{equation}
We substitute the following ansatz 
\begin{equation}
X_{i}=\lambda\Phi_{i}=\alpha \hat{G}_{i}^{(k)}
\label{matrixmodelspheresolution}
\end{equation}
into the above equation. 
The radius of a fuzzy $2k$-sphere is 
$r^{2}=\alpha^{2} n(n+2k)$ as in (\ref{radiusof2ksphere}). 
We easily find that $f_{k}$ should be given by 
\begin{eqnarray}
f_{1}=\frac{4\alpha}{\lambda}, \hspace{0.3cm}
f_{2}=\frac{4}{\alpha(n+2)}, \hspace{0.3cm}
f_{3}=\frac{3\lambda}{\alpha^{3}(n+2)(n+4)}, 
\hspace{0.3cm}
f_{4}=\frac{2\lambda^{2}}{\alpha^{5}(n+2)(n+4)(n+6)}. 
\label{explicitvalueoff}
\end{eqnarray}
Note that $f_{k}$ depends on $n$ when $k$ takes $2,3,4$. 
These $f_{k}$ are compactly denoted by 
\begin{equation}
f_{k}=-(-i)^{k}\frac{8k}{C_{k}}k!\alpha^{-2k+3}\lambda^{k-2}, 
\end{equation}
where $C_{k}$ is defined in (\ref{valueofC}). 
We can evaluate the value of the potential 
(\ref{D0matrixmodelinRR}) for the 
fuzzy $2k$-sphere solution (\ref{matrixmodelspheresolution}) as 
\begin{equation}
V_{k}=2k\left(1-\frac{4}{2k+1}\right)T_{0}
\frac{\alpha^{4}}{\lambda^{2}}
n(n+2k)N_{k}. 
\label{D0potentialvalue}
\end{equation}
It must be noted that $V_{1}$ is negative while 
$V_{k}$$(k=2,3,4)$ are positive. 

Let us comment on another classical solution. 
A set of commuting matrices 
\begin{equation}
[\Phi_{i},\Phi_{j}]=0
\label{commutingsolution}
\end{equation}
is also a classical solution. Since $\Phi_{i}$ commute 
each other, they are simultaneously 
diagonalized and represent a set of $N$ separated D$0$-branes. 
The potential energy (\ref{D0matrixmodelinRR}) 
for this solution is zero. 
Since $V_{k}$ in (\ref{D0potentialvalue}) is positive 
when $k=2,3,4$, a fuzzy $2k$-sphere is unstable and 
is expected to collapse 
into the solution (\ref{commutingsolution}). 
This situation is opposite to the case of fuzzy two-sphere. 
Since the potential energy of a fuzzy two-sphere is lower than 
that of the solution (\ref{commutingsolution}), 
a fuzzy two-sphere is classically stable 
\cite{myers}. 
This is one of differences between fuzzy two-sphere 
and fuzzy $2k$-sphere ($k\neq2$). 
Another difference can be found by noticing 
that $f_{k}$ depends on $n$ when $k$ takes $2,3,4$. 
Due to the $n$ dependence of $f_{k}$, reducible representations 
cannot be classical solutions. 
(If we require a reducible representation to be a classical 
solution of the model, an irreducible representation 
cannot. ) 
Taking account of these differences, 
the classical dynamics of higher dimensional fuzzy sphere 
is completely different from that of fuzzy two-sphere. 

A fuzzy $2k$-sphere is obtained by a matrix representation 
of the size $N_{k}$, where $N_{k}$ is defined in (\ref{sizefuzzysphere}). 
The size of matrix $N_{k}$ is interpreted as the number of D$0$-branes. 
Since the value of $N_{k}$ is restricted, the number of 
D$0$-branes is restricted to realize spherical D$2k$-branes 
from D$0$-branes. In the next section, 
$N_{k}$ is compared to the Chern number $c_{k}$ 
on spherical D$2k$-branes. 



\section{Dual description of fuzzy sphere}
\label{sec:dualdescription}
\hspace{0.4cm}
In the previous section, we realized 
a higher dimensional fuzzy sphere 
as a classical solution of a matrix model of D$0$-branes. 
Noncommutative geometry is fully used in this description. 
The classical solution is really 
a bound state of D$2k$-branes and D$0$-branes. 
On the other hand, we have a dual description. 
From the viewpoint of 
a world-volume theory on D$2k$-branes, 
D$0$-branes bounded on them 
are expressed by a nontrivial gauge field configuration. 
The Chern character of it expresses the charge of 
D$0$-branes. 
In this section, we study a dual description of the 
higher dimensional fuzzy sphere
\footnote{A dual description of fuzzy four-sphere and 
fuzzy six-sphere is discussed in \cite{hep-th/0102080,hep-th/0306250} 
in the context of a fuzzy funnel solution. 
In this paper, we focus on a relationship 
between noncommutative geometry and (nonabelian) gauge fields. 
Some parts of the calculation in these papers overlap 
with those in our paper. }. 
This study helps us to understand some unusual aspects 
of higher dimensional fuzzy spheres. 
We consider a system of spherical D$2k$-branes and 
gauge fields on them in a constant R-R $(2k+2)$-form 
field strength background, and compare the potential energy 
with (\ref{D0potentialvalue}). 
This calculation is done by using the Born-Infeld action. 
It is shown that two descriptions coincide in the limit of 
large $n$. 

Notation of indices in this section is as follows,  
$\mu,\nu=1,2,\cdots,2k+1$ and  
$\alpha,\beta=1,2,\cdots,2k$. 
$a$,$b$ are used for the world-volume indices.


\subsection{Dual description of fuzzy four-sphere}
\hspace{0.4cm}
We first consider a dual description of the fuzzy four-sphere. 
An instanton solution on a four-sphere space is constructed in 
\cite{CNYang}, and it is given 
in (\ref{SU(2)gaugefieldnorthpole}); 
\begin{equation}
A_{\alpha}=-\frac{1}{2r(r+x_{5})}
\eta_{\alpha\beta}^{i}x_{\beta}\sigma_{i}, 
\hspace{0.4cm}
A_{5}=0, 
\label{SU(2)instanton}
\end{equation}
where $\eta_{\alpha\beta}^{i}$ is the 't Hooft symbol. 
The gauge group of the instanton is $SU(2)$. 
To relate this description with fuzzy sphere, 
we need to replace $\sigma_{i}$ with $L_{i}$, where $L_{i}$ is 
the spin $n/2$ representation of $SU(2)$. 
The instanton number is defined as 
\begin{equation}
c_{2}=-\frac{1}{8\pi^{2}}\int_{S^{4}}Tr \left(F\wedge F\right).
\end{equation}
The minus sign is our convention. 
Since the instanton configuration (\ref{SU(2)instanton}) has 
the $SO(5)$ symmetry, 
we  can use the value of $F$ at the north pole 
to simplify the calculation. 
From (\ref{SU(2)fieldstrength}), 
the field strength at the north pole 
($x_{\alpha}=0$, $x_{5}=r$) is given by 
\begin{equation}
F_{\alpha\beta}=\frac{1}{2r^{2}}\eta_{\alpha\beta}^{i}L_{i}, 
\hspace{0.4cm}
F_{5\alpha}=0. 
\label{SU(2)fieldstrengthnorthpole}
\end{equation}
We now calculate the Chern numbers, which 
provide the charges of lower dimensional D-branes. 
Since $c_{1}=trF=0$, 
there are no net D$2$-brane charge. 
As can be understood from the fact that 
$L_{i}$ form a fuzzy two-sphere, 
positive charges and negative charges
cancel each other. 
It is locally nonzero 
as shown in section \ref{sec:noncommutativegaugefield}. 
$c_{2}$ is evaluated as follow, 
\begin{eqnarray}
c_{2}&=&-\frac{1}{8\pi^{2}}\frac{1}{4}\Omega_{4}r^{4}
Tr\left(\epsilon_{\alpha\beta\gamma\delta}
F_{\alpha\beta}F_{\gamma\delta}
\right) \cr
{\rule[-2mm]{0mm}{8mm}\ } 
&=&\frac{1}{16\pi^{2}}\Omega_{4}
Tr(L_{i}L_{i}) \cr
{\rule[-2mm]{0mm}{8mm}\ }  
&=&\frac{1}{6}n(n+1)(n+2) \equiv \bar{c}_{2}. 
\end{eqnarray}
This corresponds to the number of D$0$-branes. 
Therefore a bound state of $\bar{c}_{2}$ D$0$-branes
\footnote{
Due to the minus sign of $c_{2}$, it is really 
a bound state of anti-D$0$-branes and D$4$-branes.} 
and 
$(n+1)$ D$4$-branes 
is realized by introducing the $SU(2)$ gauge field. 
$F_{ab}$ in (\ref{SU(2)fieldstrengthnorthpole}) satisfies the following 
(anti-)self-dual relation,  
\begin{eqnarray}
\epsilon_{\alpha\beta\gamma\delta}F^{\gamma\delta}
=-2F_{\alpha\beta}. 
\label{selfdual}
\end{eqnarray}

Let us consider the world-volume action for $(n+1)$ D$4$-branes 
with the gauge field (\ref{SU(2)instanton}). 
The Born-Infeld action for nonabelian D$4$-branes is given by  
\begin{equation}
S_{BI}=-T_{4}\int d^{4+1}\sigma Str\sqrt{-\det(P[G+\lambda F]_{ab})}. 
\end{equation}
It is assumed that D$4$-branes are static and spherical. 
We fix the position of the D$4$-branes as 
$x_{i}=0$ ($i=6,\cdots.9$) and $x_{5}=r$, 
and adopt the static gauge $x_{a}=\sigma_{a}$ ($a=0,1,\cdots,4$) 
around the north pole. 
The pullback $P[\cdots]$ is calculated as 
\begin{eqnarray}
P[G+\lambda F]_{ab}&=&(G_{\mu\nu}+\lambda F_{\mu\nu})
\frac{\partial x^{\mu}}{\partial \sigma^{a}}
\frac{\partial x^{\nu}}{\partial \sigma^{b}} \cr
&=&G_{ab}+\lambda F_{ab}+
G_{i(a}\partial_{b)}x^{i}+ 
\lambda F_{i[a}\partial_{b]}x^{i}+
(G_{ij}+\lambda F_{ij})\partial_{a}x^{i}\partial_{b}x^{j} \cr
&\rightarrow& G_{ab}+\lambda F_{ab}. 
\end{eqnarray}
The determinant is evaluated by using these assumptions as 
\begin{eqnarray}
-\det(G_{ab}+\lambda F_{ab})
&=&1+\frac{\lambda^{2}}{2}F_{\alpha\beta}F_{\alpha\beta}
+\frac{\lambda^{4}}{64}(F_{\alpha\beta}F_{\gamma\delta}
\epsilon^{\alpha\beta\gamma\delta})^{2} \cr
{\rule[-2mm]{0mm}{10mm}\ }  
&=&\left(1+\frac{\lambda^{2}}{4}
F_{\alpha\beta}F_{\alpha\beta}\right)^{2}, 
\end{eqnarray}
where we have evaluated it around the north pole of 
a four-sphere by using the rotation symmetry, and 
a flat metric $G_{ab}=\eta_{ab}$ was used. 
From the first line to the second line, 
the self-dual condition (\ref{selfdual}) has been used. 
In this calculation, we have regard $F$ as 
commutative in spite of its nonabelian property, 
which is justified since the determinant is in 
the symmetrized trace \cite{hep-th/9701125}. 
Then the Born-Infeld action is calculated as 
\begin{eqnarray}
S&=&-T_{4}\int d^{4+1}\sigma Str\sqrt{-\det(P[G+\lambda F]_{ab})} \cr 
&=&-T_{4} \int dt
\left((n+1)
\Omega_{4} r^{4}+\bar{c}_{2}4\pi^{2}\lambda^{2}
\right)  \cr 
&=& -\int dt \left((n+1)
T_{4}\Omega_{4} r^{4}+\bar{c}_{2}T_{0}\right) , 
\label{D4BIaction}
\end{eqnarray}
where $\Omega_{4}=8\pi^{2}/3$ 
is the volume of a four-sphere with unit radius. 
The first term corresponds to the energy of 
$(n+1)$ spherical D$4$-branes with the radius $r$, 
and the second term is the energy of 
$\bar{c}_{2}$ D$0$-branes. 

We next consider the Chern-Simons coupling. 
We are now considering a case 
where only R-R six-form field strength is nonzero. 
The background R-R six-form field strength (\ref{RR2k+2}) 
can provide the following 
five form potential after we use a gauge choice, 
\begin{equation}
C^{(5)}_{t1234}=\frac{1}{5}f_{2}x_{5}\simeq \frac{1}{5}f_{2}r, 
\end{equation}
where we have evaluated around the north pole $x_{5}\simeq r$. 
Then the Chern-Simons term is calculated as 
\begin{eqnarray}
S_{CS}&=&\mu_{4}Tr\int C^{(5)} \cr 
&=&\mu_{4}Tr\int dt \int d^{4}x \left(
\frac{1}{5}f_{2}r \right ) \cr
&=& \mu_{4} \int dt  
\left((n+1)\frac{1}{5}f_{2}\Omega_{4} r^{5}\right). 
\label{D4CSaction}
\end{eqnarray}
By combining the contributions from 
the Born-Infeld action (\ref{D4BIaction}) and 
the Chern-Simons action (\ref{D4CSaction}), 
the potential energy for D$4$-branes 
is provided by 
\begin{eqnarray}
&&V(r)=\bar{c}_{2}T_{0}+
T_{4}\Omega_{4}(n+1)\left( r^{4}
-\frac{f_{2}}{5}r^{5} \right) . 
\end{eqnarray}
We regard this potential as a function of $r$. 
It has two extrema, one is given by $r=0$ and 
another is 
\begin{equation}
r=\frac{4}{f_{2}}=\alpha (n+2) \equiv r_{\star}. 
\end{equation}
At the first extremum, D$4$-branes cannot have a nonzero radius 
and only $\bar{c}_{2}$ D$0$-branes can exist. 
The second one is related to a fuzzy four-sphere solution. 
$r_{\star}$ is the radius of the four-sphere, and 
can be compared to the radius of a fuzzy four sphere 
which is given in (\ref{radiusof2ksphere}). 
Because both becomes $\alpha n$ at large $n$, two descriptions provide 
the same radius at large $n$. 

From the shape of the potential, the first extremum is 
a local minimum, while the second one is a local maximum. 
Therefore the spherical configuration is classically unstable 
against a small fluctuation. 
This situation is the same as one encountered in 
the previous section. 

We now calculate the value of the potential 
for the spherical solution. 
By substituting $r=r_{\star}$ into $V(r)$, 
we have 
\begin{equation}
V(r_{\star})-\bar{c}_{2}T_{0}=T_{0}\frac{2}{15}
\frac{\alpha^{4}}{\lambda^{2}}(n+1)(n+2)^{4}
\simeq 
T_{0}\frac{2}{15}
\frac{\alpha^{4}}{\lambda^{2}}n^{5}. 
\end{equation}
This should be compared with $V_{2}$, 
which is obtained in (\ref{D0potentialvalue});
\begin{equation}
V_{2}=T_{0}\frac{2}{15}
\frac{\alpha^{4}}{\lambda^{2}}
n(n+1)(n+2)(n+3)(n+4) \simeq 
T_{0}\frac{2}{15}
\frac{\alpha^{4}}{\lambda^{2}}n^{5}. 
\end{equation}
These two values agree at large $n$. 
We can also compare the D$0$-brane charge. 
In the first description, 
the size of the matrix represents the D$0$-brane charge 
and it is given by $N_{2}$. 
On the other hand, it is $\bar{c}_{2}$ in the second 
description. Both behave $n^{3}/6$ at large $n$. 
We have compared three quantities, the potential energy, 
the radius and the D$0$-brane charge. 
All of them gave the same values in two descriptions when 
$n$ is large. 
This result leads to the conclusion that 
a fuzzy four-sphere is the same object as 
spherical D$4$-branes with an $SU(2)$ monopole gauge field 
at large $n$. 

\vspace{0.2cm}

Let us comment on the validity of 
these two descriptions 
\cite{hep-th/0102080,hep-th/0303072}. 
We first considered the world-volume theory of D$0$-branes. 
The assumption $\lambda [\Phi_{i},\Phi_{j}] \ll 1$ 
was used to derive the low energy effective action 
of D$0$-branes (\ref{BIpart}). 
Since this condition is rewritten as 
\begin{equation}
l_{nc}^{2} \simeq \frac{r^{2}}{n} \ll l_{s}^{2}, 
\label{validityD0}
\end{equation} 
the noncommutative scale $l_{nc}$, 
which is defined in (\ref{noncommutativescale}), 
has to be much smaller than the string scale $l_{s}$. 
On the other hand, the computations by using 
the world-volume action of D$2k$-branes can be trusted  
as long as the field strength is slowly varying 
$\mid l_{s}\partial F\mid \ll \mid F\mid$. 
This condition is satisfied when the radius 
is much larger than the string scale;  
\begin{equation}
l_{s} \ll r. 
\label{validityD2k}
\end{equation}
If $r$ satisfies the following region 
\begin{equation}
l_{s} \ll r \ll \sqrt{n}l_{s}, 
\label{validregion}
\end{equation}
both of (\ref{validityD0}) and (\ref{validityD2k}) 
are satisfied. 
It is sometimes convenient to rewrite (\ref{validregion}) 
as 
\begin{equation}
\frac{l_{s}}{\sqrt{n}} \ll l_{nc} \ll l_{s}.
\label{validregion2}
\end{equation}
We can expect the agreement of two descriptions 
in a large region 
by taking a large $n$ limit. 
This is the reason why we could obtain the agreement 
of two descriptions in a large $n$ limit. 
We can provide another explanation 
in section \ref{sec:noncommutativegaugefield}. 


\subsection{Dual description of fuzzy six-sphere}
\label{sec;dualsixsphere}
\hspace{0.4cm}
We next consider a dual description of a fuzzy six-sphere. 
The idea is basically the same as the case of 
fuzzy four-sphere. 
 
We begin by introducing an $SO(6)$ gauge field 
on a six-sphere which is obtained in (\ref{spin2kgaugefield}): 
\begin{eqnarray}
A_{\alpha}=\frac{-i}{2r(r+x_{2k+1})}
\Sigma_{\beta\alpha}^{N}x_{\beta}, 
\hspace{0.4cm}
A_{7}=0
\label{spin6gaugefield}
\end{eqnarray}
where 
$\Sigma_{\alpha\beta}^{N}=(\Sigma_{ij}^{N},\Sigma_{6i}^{N})
=(\gamma_{ij},i\gamma_{i})
=(\frac{1}{2}[\gamma_{i},\gamma_{j}],i\gamma_{i})$, 
$\gamma_{i}$ $(i=1,\cdots,5)$ is the five-dimensional gamma matrix, 
whose explicit representation is provided 
in (\ref{fivedimensionalgammamatrix}). 
$\Sigma_{\alpha\beta}^{N}$ transforms in 
a spinor representation of $SO(6)$. 
To make a connection to fuzzy sphere, 
we need to consider a higher dimensional representation of $SO(6)$. 
Such a representation was already obtained 
in section \ref{sec:fuzzysphere}, and  
the $k=2$ case is relevant in the present case. 
Then we replace the five-dimensional gamma matrices 
$\gamma_{i}$ and $\gamma_{ij}$ with 
higher dimensional representations 
$\hat{G}_{i}^{(2)}$ and $\hat{G}_{ij}^{(2)}$. 
The index $(2)$ was added to emphasize that 
the matrices are related to the $k=2$ case 
in section \ref{sec:fuzzysphere}. 
Accordingly $\Sigma_{\alpha\beta}^{N}$ is replaced with 
$\hat{\Sigma}_{\alpha\beta}^{N}=
(\hat{\Sigma}_{ij}^{N},\hat{\Sigma}_{2k,i}^{N})
\equiv(\hat{G}_{ij}^{(2)},i\hat{G}_{i}^{(2)})$. 
Since these matrices are realized by the size $N_{2}$ 
which is defined in (\ref{sizefuzzysphere}), 
the gauge field becomes an $N_{2}\times N_{2}$ matrix. 
A interesting fact is that these matrices 
construct a fuzzy four-sphere. 

Before we begin calculations, let us explain 
the meaning of the index $N$ in $\hat{\Sigma}_{\alpha\beta}^{N}$. 
We consider the $N_{3}$-dimensional 
irreducible representation 
of $SO(7)$, which is denoted by $\hat{G}_{\mu\nu}^{(3)}$ 
$(\mu,\nu=1,\cdots,7)$. 
As is explained in section \ref{sec:fuzzysphere}, 
it is associated with a fuzzy six-sphere. 
We now restrict our interest to a subalgebra 
$\hat{G}_{\alpha\beta}^{(3)}$ $(\alpha,\beta=1,\cdots,6)$, 
which forms the $SO(6)$ algebra. 
$\hat{G}_{\alpha\beta}^{(3)}$ is reducible, 
and is characterized by the eigenvalues of 
$\hat{G}_{7}^{(3)}=diag(n,n-2,\cdots,-n+2-n)$. 
$\hat{G}_{7}^{(3)}$ is a generalized chirality matrix. 
We consider a subspace which is labelled by $\hat{G}_{7}^{(3)}=n1$, 
which corresponds to the $N$orth pole of a fuzzy six-sphere. 
It can be shown that the size of the unit matrix 1 
is $N_{2}$. 
This $\hat{G}_{\alpha\beta}^{(3)}$ which is labelled by 
$\hat{G}_{7}^{(3)}=n1$ is nothing 
but $\hat{\Sigma}_{\alpha\beta}^{N}$. 
We added the index $N$ to clarify a relationship 
between the fuzzy six-sphere and the $SO(6)$ 
nonabelian gauge field. 

The field strength at the north pole 
($x_{\alpha}=0$, $x_{7}=r$), 
which is calculated in 
(\ref{fieldstrengthspin(2k+1)}), is 
\begin{equation}
F_{\alpha\beta}=-\frac{i}{2r^{2}}\hat{\Sigma}_{\alpha\beta}^{N},
\hspace{0.4cm}
F_{7\alpha}=0. 
\label{fieldstrengthspin4}
\end{equation}
The calculation of the Chern numbers 
is easily done by making full use of 
some formulae in appendix \ref{sec:someformulaefuzzysphere}. 
The first and second Chern numbers are 
calculated as 
$c_{1}=TrF \simeq Tr\Sigma_{\alpha\beta}=0$ and 
\begin{eqnarray}
c_{2}&\sim&Tr\left(F\wedge F\right) \cr
&\sim&
\epsilon_{\alpha_{1}\alpha_{2}\alpha_{3}\alpha_{4}}
Tr\left(F_{\alpha_{1}\alpha_{2}}F_{\alpha_{3}\alpha_{4}}
\right) \cr 
&\sim& \epsilon_{i_{1}i_{2}i_{3}i_{4}}
Tr\left(\hat{G}_{i_{1}i_{2}}^{(2)}\hat{G}_{i_{3}i_{4}}^{(2)}\right)
+\epsilon_{i_{1}i_{2}i_{3}}
Tr\left(\hat{G}_{i_{1}i_{2}}^{(2)}\hat{G}_{i_{3}}^{(2)}\right) \cr
&\sim&
Tr\left(\hat{G}_{i}^{(2)}\right)
+Tr\left(\hat{G}_{ij}^{(2)}\right)=0. 
\label{c2=0onD6}
\end{eqnarray}
The net charge of 
D$2$-branes and that of D$4$-branes vanish. 
This is equivalent to the fact that 
the net charge of a fuzzy two-sphere and 
that of a fuzzy four-sphere are zero. 
The third Chern number is 
\begin{eqnarray}
c_{3}&=&\frac{1}{48\pi^{3}}\int_{S^{6}}
Tr\left( F\wedge F \wedge F \right)\cr 
&=&\frac{1}{48\pi^{3}}\frac{1}{8}
\Omega_{6}r^{6}
\sum_{\alpha_{i}=1}^{6}
Tr\left(
F_{\alpha_{1}\alpha_{2}}
F_{\alpha_{3}\alpha_{4}}
F_{\alpha_{5}\alpha_{6}}\right) 
\epsilon_{\alpha_{1}\alpha_{2}\alpha_{3}\alpha_{4}
\alpha_{5}\alpha_{6}}
\cr
&=&\frac{1}{512\pi^{3}}\Omega_{6}
\sum_{i_{i}=1}^{5}
Tr\left(
\hat{G}_{i_{1}}^{(2)}\hat{G}_{i_{2}}^{(2)}\hat{G}_{i_{3}}^{(2)}
\hat{G}_{i_{4}}^{(2)}\hat{G}_{i_{5}}^{(2)}
\right)\epsilon_{i_{1}i_{2}i_{3}i_{4}i_{5}} \cr 
&=&\frac{1}{512\pi^{3}}\Omega_{6}
(8n+16)Tr(\hat{G}_{i}^{(2)}\hat{G}_{i}^{(2)}) \cr 
{\rule[-2mm]{0mm}{8mm}\ }  
&=&\frac{1}{360}n(n+1)(n+2)^{2}(n+3)(n+4)
\equiv \bar{c}_{3}. 
\end{eqnarray}
$\Omega_{6}=16\pi^{3}/15$ is the volume of a six-sphere with unit radius. 
This corresponds to the number of D$0$-branes, 
and can be compared with $N_{3}$ in (\ref{sizefuzzysphere}). 
We easily find that these coincide at large $n$. 
The use of the nonabelian gauge field (\ref{spin6gaugefield}) 
allows us to construct a bound state of $\bar{c}_{3}$ 
D$0$-branes and $N_{2}$ D$6$-branes. 
We note that (\ref{fieldstrengthspin4}) satisfies 
\begin{eqnarray}
\epsilon_{\alpha_{1}\alpha_{2}\alpha_{3}\alpha_{4}\alpha_{5}\alpha_{6}
}F^{\alpha_{3}\alpha_{4}}
F^{\alpha_{5}\alpha_{6}}=
\frac{4}{r^{2}}(n+2)F_{\alpha_{1}\alpha_{2}},  
\end{eqnarray}
which is a natural generalization of a self-dual equation of 
the instanton. 

We now consider a world-volume action for 
$N_{2}$ $D6$-branes with the gauge field 
(\ref{fieldstrengthspin4}). The Born-Infeld action is 
\begin{equation}
S_{BI}=-T_{6}\int d^{6+1}\sigma 
Str\sqrt{-\det(P[G+\lambda F]_{ab})}.  
\end{equation} 
The determinant is evaluated as follows, 
\begin{eqnarray}
-\det(G_{ab}+\lambda F_{ab})
&=&1+\frac{\lambda^{2}}{2}F_{\alpha\beta}F_{\alpha\beta}
\cr 
{\rule[-2mm]{0mm}{8mm}\ } 
&&
+\frac{\lambda^{4}}{128}(F_{\alpha_{1}\alpha_{2}}F_{\alpha_{3}\alpha_{4}}
\epsilon^{\alpha\beta\alpha_{1}\alpha_{2}\alpha_{3}\alpha_{4}}) 
(F_{\beta_{1}\beta_{2}}F_{\beta_{3}\beta_{4}}
\epsilon^{\alpha\beta\beta_{1}\beta_{2}\beta_{3}\beta_{4}}) 
\cr 
{\rule[-2mm]{0mm}{8mm}\ } 
\hspace{3cm}
&&+\frac{\lambda^{6}}{2304}
\left(\epsilon^{\alpha_{1}\alpha_{2}\alpha_{3}\alpha_{4}
\alpha_{5}\alpha_{6}}
F_{\alpha_{1}\alpha_{2}}F_{\alpha_{3}\alpha_{4}}F_{\alpha_{5}\alpha_{6}}
\right)^{2} \cr 
{\rule[-2mm]{0mm}{8mm}\ } 
&=&
1+\frac{3}{4}\left(\frac{\lambda}{r^{2}}\right)^{2}n(n+6)
+\frac{3}{16}\left(\frac{\lambda}{r^{2}}\right)^{4}n(n+2)^{2}(n+6) 
{\rule[-2mm]{0mm}{8mm}\ } 
\cr 
&&+\frac{1}{64}\left(\frac{\lambda}{r^{2}}\right)^{6}
n^{2}(n+2)^{2}(n+4)^{2} \cr 
{\rule[-2mm]{0mm}{8mm}\ } 
&\simeq& 
1+\frac{3}{4}\left(\frac{\lambda n}{r^{2}}\right)^{2}
+\frac{3}{16}\left(\frac{\lambda n}{r^{2}}\right)^{4} 
+\frac{1}{64}\left(\frac{\lambda n}{r^{2}}\right)^{6}, 
\end{eqnarray}
where we have assumed a large $n$ to obtain the last 
expression. 
We define $y\equiv r^{2}/\lambda n$. 
The Born-Infeld action becomes
\begin{eqnarray}
S_{BI}/\int dt
&=&-T_{6}\Omega_{6}r^{6}N_{2} 
\sqrt{
1+\frac{3}{4y^{2}}
+\frac{3}{16y^{4}}+\frac{1}{64y^{6}}
} \cr
{\rule[-2mm]{0mm}{8mm}\ } 
&=&-T_{6}\Omega_{6}r^{6}N_{2} 
\frac{1}{8y^{3}}
\sqrt{1+12y^{2}+48y^{4}+64y^{6}}\cr
{\rule[-2mm]{0mm}{8mm}\ } 
&\simeq&-T_{6}\Omega_{6}r^{6}N_{2} 
\left(\frac{1}{8y^{3}}+\frac{3}{4y}+
O(y)\right) \cr
{\rule[-2mm]{0mm}{8mm}\ } 
&\simeq&
-\bar{c}_{3}T_{0}
-\frac{3}{4}T_{6}N_{2}\Omega_{6}r^{4}
\lambda n . 
\end{eqnarray}
To arrive the last expression, we have expanded the square root by 
assuming the condition $ y \ll 1$. 
The use of this assumption is valid since it satisfies 
(\ref{validregion}). 
We next consider the coupling of D$6$-branes 
to the external R-R field. 
The constant R-R eight-form field strength 
background (\ref{RR2k+2}) provides 
the following R-R seven-form field in a certain gauge, 
\begin{equation}
C^{(7)}_{t123456}=\frac{1}{7}f_{3}x_{7}\simeq \frac{1}{7}f_{3}r.
\end{equation}
The Chern-Simons term is calculated as
\begin{eqnarray}
S_{CS}
= \mu_{6} \int dt
\left(N_{2}\frac{f_{3}}{7}\Omega_{6}r^{7}\right), 
\end{eqnarray}
and the potential for the D$6$-branes is provided by 
\begin{equation}
V(r)=\bar{c}_{3}T_{0}+T_{6}\Omega_{6}N_{2}
\left( \frac{3}{4}\lambda n r^{4}
-\frac{f_{3}}{7}r^{7}\right). 
\end{equation}
The first term represents the rest energy of $\bar{c}_{3}$ 
D$0$-branes. We search for extrema of this potential 
by regarding it as a function of $r$. 
We can find two extrema, one is trivial extremum $r=0$, 
and another is 
\begin{equation}
r=\sqrt[3]{\frac{3\lambda n}{f_{3}}}
=\sqrt[3]{\alpha^{3}n(n+2)(n+4)}
\simeq \alpha n 
\equiv r_{\star}.
\end{equation}
This extremum corresponds to a fuzzy six-sphere solution 
in the matrix model of D$0$-branes. 
The radius of this spherical solution agrees with 
that of the fuzzy six sphere (\ref{radiusof2ksphere}) 
at large $n$. 
The potential value for this extremum is found to be 
\begin{equation}
V(r_{\star})-\bar{c}_{3}T_{0}
\simeq 
T_{0}\frac{1}{140}
\frac{\alpha^{4}}{\lambda^{2}}n^{8}. 
\end{equation}
We recall the D$0$-brane calculation in the previous section; 
\begin{eqnarray}
V_{3}&=&T_{0}\frac{1}{140}
\frac{\alpha^{4}}{\lambda^{2}}
n(n+1)(n+2)(n+3)^{2}(n+4)(n+5)(n+6) \cr
&\simeq& T_{0}\frac{1}{140}
\frac{\alpha^{4}}{\lambda^{2}}n^{8}. 
\end{eqnarray}
$V(r_{\star})$ and $V_{3}$ give the same value 
including the coefficient when $n$ is large. 
These results manifest the fact that 
a fuzzy six-sphere of the matrix size $N_{3}$ is 
the same object as 
$N_{2}$ spherical D$6$-branes with a nonabelian 
$SO(6)$ gauge field.


\subsection{Dual description of fuzzy eight-sphere}
\hspace{0.4cm}
It is straightforward to generalize the calculations in the previous 
two cases to a dual description of a fuzzy eight-sphere.  
The detailed calculations are almost analogous to 
the previous cases. 
We start with an $SO(8)$ monopole field on an eight-sphere; 
\begin{eqnarray}
A_{\alpha}=\frac{-i}{2r(r+x_{2k+1})}
\Sigma_{\beta\alpha}^{N}x_{\beta}, 
\hspace{0.4cm}
A_{9}=0
\label{gaugefieldspin8}
\end{eqnarray}
where $\Sigma_{\alpha\beta}^{N}=
(\Sigma_{ij}^{N},\Sigma_{8i}^{N})
=(\gamma_{ij},i\gamma_{i})$, 
and $\gamma_{i}$ $(i=1,\cdots,7)$ 
is the seven-dimensional gamma matrix. 
The gauge field belongs to 
a spinor representation of $SO(8)$. 
To construct a higher dimensional representation, 
we replace $\Sigma_{\alpha\beta}^{N}$ 
with $\hat{\Sigma}_{\alpha\beta}^{N} \equiv 
(\hat{G}_{ij}^{(3)},i\hat{G}_{i}^{(3)})$. 
The index (3) means that 
these are the matrices of the $k=3$ case 
in section \ref{sec:fuzzysphere}. 
These are realized by an $N_{3}\times N_{3}$ matrices, 
where $N_{3}$ is presented in (\ref{sizefuzzysphere}). 
Note that $\hat{G}_{ij}^{(3)}$ and $\hat{G}_{i}^{(3)}$ 
construct a fuzzy six-sphere. 

The gauge field strength is calculated 
in (\ref{fieldstrengthspin(2k+1)}), and it becomes 
the following form at the north pole 
($x_{\alpha}=0$, $x_{9}=r$); 
\begin{equation}
F_{\alpha\beta}=-\frac{i}{2r^{2}}\hat{\Sigma}_{\alpha\beta}^{N},
\hspace{0.4cm}
F_{9\alpha}=0. 
\label{spin(8)fieldstrengthnorthpole}
\end{equation}
The vanishing of the Chern numbers 
$c_{1}=c_{2}=c_{3}=0$ is shown 
by the analogous calculations to (\ref{c2=0onD6}) 
with the help of some properties of $\hat{G}_{i}^{(3)}$ and 
$\hat{G}_{ij}^{(3)}$. 
Therefore the net charge of D$2k^{\prime}$-brane 
($k^{\prime}=1,2,3$), 
which forms a fuzzy $2k^{\prime}$-sphere, is zero. 
The forth Chern number is nonzero; 
\begin{eqnarray}
c_{4}&=&-\frac{1}{4!(2\pi)^{4}}\int_{S^{8}}
Tr \left(F\wedge F \wedge F \wedge F\right) \cr 
&=&
-\frac{1}{4!(2\pi)^{4}}
\frac{1}{2^{4}}\Omega_{8}r^{8}
\sum_{\alpha_{i}=1}^{8}
Tr\left(F_{\alpha_{1}\alpha_{2}}
F_{\alpha_{3}\alpha_{4}}F_{\alpha_{5}\alpha_{6}}
F_{\alpha_{7}\alpha_{8}}\right) 
\epsilon_{
\alpha_{1}\alpha_{2}\alpha_{3}\alpha_{4}
\alpha_{5}\alpha_{6}\alpha_{7}\alpha_{8}
}\cr 
&=&
\frac{i}{4!(2\pi)^{4}}\frac{1}{2^{5}}\Omega_{8}
\sum_{i_{i}=1}^{7}
Tr\left(
\hat{G}_{i_{1}}^{(3)}\hat{G}_{i_{2}}^{(3)}
\hat{G}_{i_{3}}^{(3)}\hat{G}_{i_{4}}^{(3)}
\hat{G}_{i_{5}}^{(3)}\hat{G}_{i_{6}}^{(3)}
\hat{G}_{i_{7}}^{(3)}
\right)\epsilon_{i_{1}i_{2}i_{3}i_{4}i_{5}i_{6}i_{7}} \cr
&=&
\frac{1}{302400}n(n+1)(n+2)^{2}(n+3)^{2}(n+4)^{2}(n+5)(n+6)
\equiv \bar{c}_{4}. 
\end{eqnarray}
$\Omega_{8}=32\pi^{4}/105$ 
is the volume of an eight-sphere with unit radius. 
This represents the number of D$0$-branes. 
We can confirm the agreement with $N_{4}$ at large $n$. 
We note that (\ref{spin(8)fieldstrengthnorthpole}) 
satisfies 
\begin{eqnarray}
\epsilon_{\alpha_{1}\alpha_{2}\alpha_{3}\alpha_{4}\alpha_{5}
\alpha_{6}\alpha_{7}\alpha_{8}
}F^{\alpha_{3}\alpha_{4}}F^{\alpha_{5}\alpha_{6}}
F^{\alpha_{7}\alpha_{8}}=
-\frac{12}{r^{4}}(n+2)(n+4)F_{\alpha_{1}\alpha_{2}}.  
\label{selfdual8sphere}
\end{eqnarray}

Let us next consider a world-volume theory on D$8$-branes. 
The dynamics of D$8$-branes with a gauge field 
is described by the Born-Infeld action; 
\begin{equation}
S_{BI}=-T_{8}\int d^{8+1}\sigma 
Str\sqrt{-\det(P[G+\lambda F]_{ab})}. 
\end{equation} 
The determinant around the north pole becomes 
\begin{eqnarray}
-\det(G_{ab}+\lambda F_{ab})
&=&1+\frac{\lambda^{2}}{2}F_{\alpha\beta}F_{\alpha\beta} \cr
&&\hspace{-2cm}
+\frac{\lambda^{4}}{1536}(F_{\alpha_{1}\alpha_{2}}
F_{\alpha_{3}\alpha_{4}}
\epsilon^{\alpha\beta\gamma\delta
\alpha_{1}\alpha_{2}\alpha_{3}\alpha_{4}}) 
(F_{\beta_{1}\beta_{2}}F_{\beta_{3}\beta_{4}}
\epsilon^{\alpha\beta\gamma\delta\beta_{1}\beta_{2}\beta_{3}\beta_{4}}) 
\cr 
&&\hspace{-2cm}
+\frac{\lambda^{6}}{4608}
\left(\epsilon^{\alpha\beta\alpha_{1}\alpha_{2}\alpha_{3}
\alpha_{4}\alpha_{5}\alpha_{6}}
F_{\alpha_{1}\alpha_{2}}F_{\alpha_{3}\alpha_{4}}
F_{\alpha_{5}\alpha_{6}}
\right)
\left(\epsilon^{\alpha\beta\beta_{1}\beta_{2}\beta_{3}\beta_{4}
\beta_{5}\beta_{6}}
F_{\beta_{1}\beta_{2}}F_{\beta_{3}\beta_{4}}
F_{\beta_{5}\beta_{6}}
\right) \cr 
&&\hspace{-2cm}
+\frac{\lambda^{8}}{147456}
\left(\epsilon^{\alpha_{1}\alpha_{2}\alpha_{3}\alpha_{4}
\alpha_{5}\alpha_{6}\alpha_{7}\alpha_{8}}
F_{\alpha_{1}\alpha_{2}}F_{\alpha_{3}\alpha_{4}}
F_{\alpha_{5}\alpha_{6}}F_{\alpha_{7}\alpha_{8}}
\right)^{2}. 
\end{eqnarray}
By assuming the condition $r^{2} \ll \lambda n$, 
and using (\ref{selfdual8sphere}) 
and some formulae in appendix \ref{sec:someformulaefuzzysphere}, 
the resulting action is 
\begin{equation}
S_{BI}\simeq \int dt \left(-\bar{c}_{4}T_{0}
-\frac{1}{2}T_{8}N_{3}\Omega_{8}r^{4}
\lambda^{2}n^{2}
\right). 
\end{equation}
The Chern-Simons term is also calculated as 
\begin{equation}
S_{CS}=\mu_{8}\int dt \left(
N_{3}\frac{f_{4}}{9}\Omega_{8}r^{9}\right), 
\end{equation}
where we have used the fact that 
the R-R field is given by the following form 
in a certain gauge, 
\begin{equation}
C^{(9)}_{t12345678}=\frac{1}{9}f_{4}x_{9}\simeq \frac{1}{9}f_{4}r.
\end{equation}
Therefore 
the dynamics of static D$8$-branes with the gauge field 
in a constant R-R field strength background 
is described by
\begin{eqnarray}
V(r)=\bar{c}_{4}T_{0}+T_{8}\Omega_{8}N_{3}
\left(\frac{1}{2}r^{4}
\lambda^{2} n^{2}
-\frac{f_{4}}{9}r^{9}\right).  
\end{eqnarray}
This potential indicates two extrema, $r=0$ and 
\begin{equation}
r=\sqrt[5]{
\frac{2\lambda^{2} n^{2}}{f_{4}}}
\simeq \alpha n 
\equiv r_{\star}.
\end{equation}
The second one can be compared with the radius of a fuzzy eight-sphere. 
$r_{\star}$ surely agrees with the 
radius of a fuzzy eight-sphere at large $n$. 
The potential value for $r=r_{\star}$ is 
\begin{equation}
V(r_{\star})-\bar{c}_{4}T_{0}
\simeq 
T_{0}\frac{1}{68040}
\frac{\alpha^{4}}{\lambda^{2}}n^{12}. 
\end{equation}
This presents the same large $n$ behavior as $V_{4}$ 
including the numerical coefficient. 
One can state that 
a fuzzy eight-sphere is the same object as 
D$8$-branes with the nonabelian gauge field 
(\ref{gaugefieldspin8}) when $n$ is large.


\section{Noncommutativity and nonabelian gauge field}
\label{sec:noncommutativegaugefield}
\hspace{0.4cm}
In the previous sections, we considered two world-volume theories. 
Since they provide the same values for various 
quantities at large $n$, 
they are supposed to be the same thing. 
In this section, we discuss a concrete correspondence of them 
by noticing a relationship 
between the noncommutativity of fuzzy sphere 
and nonabelian gauge fields. 
The results presented in the previous section 
confirmed that a fuzzy $2k$-sphere is dual to 
D$2k$-branes with an $SO(2k)$ nonabelian gauge field. 
An interesting fact is that the nonabelian gauge field 
on D$2k$-branes is expressed by 
matrices which are related to the fuzzy $(2k-2)$-sphere. 
We recall the gauge field strength evaluated 
at the north pole ($x_{\alpha}=0$, $x_{2k+1}=r$); 
\begin{equation}
F_{\alpha\beta}=-\frac{i}{2r^{2}}
\hat{\Sigma}_{\alpha\beta}^{N}, 
\label{fieldstrengthgeneral}
\end{equation}
where 
$\hat{\Sigma}_{\alpha\beta}^{N}=
(\hat{\Sigma}_{ij}^{N},\hat{\Sigma}_{2k,i}^{N})
=(\hat{G}_{ij}^{(k-1)},i\hat{G}_{i}^{(k-1)})$ 
$(i=1,\cdots,2k-1)$. 
The matrices $\hat{G}_{ij}^{(k-1)}$ and $\hat{G}_{i}^{(k-1)}$ 
are realized by the $N_{k-1}$-dimensional irreducible 
representation of $SO(2k-1)$, forming 
the fuzzy $(2k-2)$-sphere algebra. 


A commutation relation of coordinates of a $2k$-fuzzy sphere 
is given by the first equation in (\ref{fuzzyspherealgebra2}). 
This relation reduces to the following relation 
at the north pole of a fuzzy $2k$-sphere ($\hat{G}_{2k+1}=n$), 
\begin{equation}
[\hat{G}_{\alpha}^{N},\hat{G}_{\beta}^{N}] 
=2\hat{G}_{\alpha\beta}^{N}, 
\label{commutationrelationinnorthpole}
\end{equation} 
where $\alpha,\beta=1,\cdots,2k$. 
We should not confuse two kinds of the north pole. 
As is explained in \ref{sec;dualsixsphere}, 
$\hat{G}_{\alpha\beta}^{N}$ is 
the same matrix as $\Sigma_{\alpha\beta}^{N}$. 
Both are realized by an $N_{k-1}$-dimensional representation of 
$SO(2k)$. Since the two descriptions 
agree in the large $n$ limit, 
we may combine two relations (\ref{fieldstrengthgeneral}) and 
(\ref{commutationrelationinnorthpole}) 
only in the large $n$ limit. 
Therefore we are led to the following noncommutative relation for 
coordinates of a sphere, 
\begin{equation}
[X_{\alpha},X_{\beta}]=4i\alpha^{2}r^{2}F_{\alpha\beta}, 
\label{commutationfieldstrengthnorth}
\end{equation}
where $X_{\alpha}=\alpha \hat{G}_{\alpha}^{N}$. 
We have identified the noncommutative coordinates 
with commutative coordinates in the dual description, and 
the noncommutativity of coordinates has been given by 
the field strength. 
This relation suggests the identification of 
the noncommutative coordinate with the covariant derivative 
under the $SO(2k)$ monopole background. 
This is also expected from the result in \cite{hep-th/0204256}, 
where it was shown that an adjoint action of $\hat{G}_{\mu}$ 
is mapped to 
the covariant derivative under the monopole background for 
the $k=2$ case 
\footnote{ 
Functions on fuzzy spheres are expanded by 
an extended version of spherical harmonics \cite{hep-th/0105006}. 
The spectrum of $ad(G)^{2}$ was calculated by acting 
on the spherical harmonics, and compared it to 
the Hamiltonian in the four-dimensional quantum Hall system, 
which is the square of the covariant derivative.}.

The commutation relation (\ref{commutationfieldstrengthnorth}) 
is valid at the north pole. It is natural to expect 
that this relation exists 
without restricting to the north pole 
because a (fuzzy) $2k$-sphere has the $SO(2k+1)$ symmetry. 
Therefore we suppose 
\begin{equation}
[X_{\mu},X_{\nu}]
=2\alpha^{2}\hat{G}_{\mu\nu}
=4i\alpha^{2}r^{2}F_{\mu\nu}, 
\label{identifyGwithF}
\end{equation}
where the size of the matrices is $N_{k}$ 
and $\mu,\nu=1,\cdots,2k+1$. 
$F_{\mu\nu}$ has been expressed by the 
$N_{k}$-dimensional irreducible representation of $SO(2k+1)$. 
This relation is interesting in the following sense. 
In the fuzzy sphere algebra 
(\ref{so(2k+1)algebra}) and (\ref{fuzzyspherealgebra2}), 
$\hat{G}_{\mu\nu}$ and $\hat{G}_{\mu}$ are treated on the same footing. 
If we regard $\hat{G}_{\mu}$ as coordinates of a fuzzy sphere, 
$\hat{G}_{\mu\nu}$ acts on them as the $SO(2k+1)$ rotation generator 
(we may regard $\hat{G}_{\mu\nu}$ as coordinates of 
an extra space.).  
On the other hand, the role of coordinates 
and that of a field strength in the dual description 
are clearly different. 
The relation (\ref{identifyGwithF}) suggests that 
the internal degrees of freedom $F_{\mu\nu}$ 
should be identified with the rotation generator. 
This is one of the characteristic aspects of 
noncommutative geometry. 
Such identification can be seen as 
the lowest Landau level physics 
in a higher dimensional quantum Hall system 
\cite{zhanghu,cond-mat/0306045,hep-th/0310274}
\footnote{
Some works investigating higher dimensional 
quantum Hall system 
have also been reported in \cite{
hep-th/0201016,hep-th/0203095,cond-mat/0206164,
hep-th/0209182,cond-mat/0306351}. }. 
It is obtained by considering a motion of 
electrons in a monopole gauge field background \cite{CNYang}. 
Considering the analogy to the quantum Hall system 
can provide an intuitive explanation 
for the agreement of two descriptions at large $n$. 
In this system, the angular momentum operator 
$\Lambda_{\mu\nu}=-i(x_{\mu}D_{\nu}-x_{\nu}D_{\mu})$ 
and the field strength (\ref{fieldstrengthspin(2k+1)}) 
have the following relation; 
\begin{equation}
G_{\mu\nu}=\Lambda_{\mu\nu}+2ir^{2}F_{\mu\nu}. 
\label{relationofqhsystem}
\end{equation}
$\Lambda_{\mu\nu}$ does not satisfy the $SO(2k+1)$ algebra 
due to the existence of the monopole background, and 
it is $G_{\mu\nu}$ that satisfies 
the $SO(2k+1)$ algebra. 
The angular momentum generated by $\Lambda_{\mu\nu}$ 
characterizes the Landau level, and therefore 
the representation of $G_{\mu\nu}$ depends on the Landau level. 
The restriction to the lowest Landau level is 
achieved by $\Lambda_{\mu\nu}\simeq 0$. 
Since the magnitude of the field strength $F$ in 
(\ref{fieldstrengthgeneral}) is given by 
$O(n/r^{2})$, the contribution of 
the second term in the right hand side of (\ref{relationofqhsystem}) 
becomes large compared to the first term 
in a large $n$ limit. 
Therefore the field strength $F_{\mu\nu}$ is identified 
with the rotation generator $G_{\mu\nu}$ in the lowest Landau level. 
It can be also shown that $G_{\mu\nu}$ is given by 
the spinor representation of $SO(2k+1)$ 
\cite{zhanghu,cond-mat/0306045,hep-th/0310274}. 
This limit is just the strong magnetic field limit. 
As is well known in the two-dimensional quantum Hall system, 
guiding center coordinates are identified with 
coordinates of electrons 
in the strong magnetic field limit. 
Accordingly coordinates of electrons are described by 
noncommutative geometry. 
Fuzzy spheres are actually realized in the higher dimensional 
quantum Hall system after we take the lowest Landau level limit. 
This is an intuitive explanation for the agreement of 
two descriptions in the large $n$ limit. 

In the previous paragraph, we have discussed a large $n$ limit. 
We have two parameters $r$ and $l_{nc}$, 
and the ratio between them is $r/l_{nc}\simeq \sqrt{n}$. 
Therefore there are two kinds of large $n$ limit . 
One is realized by taking a large $r$ limit 
with keeping the noncommutative scale $l_{nc}$, 
the other by taking a small $l_{nc}$ limit 
with keeping the radius $r$. 
Note that the radius $r$ is not a physical radius of the sphere. 
These two limits have the following meanings. 
If we see a fuzzy sphere from a short distance, 
a noncommutative structure can be seen. 
On the other hand, if we see from a long 
distance, a fuzzy sphere looks like a commutative sphere. 
These parameters have to 
satisfy (\ref{validregion}) or (\ref{validregion2}). 
In the first limit, we should consider (\ref{validregion2}). 
We need to take a large $n$ limit with 
keeping $l_{nc}$ small compared to the string scale. 
In the second limit, 
we need to take a large $n$ limit with 
keeping $r$ large compared to the string scale. 
The first limit is the same as one which is used in \cite{SW}. 

We shall now look more carefully into 
the equation (\ref{identifyGwithF}). 
In (\ref{identifyGwithF}), the field strength was regarded 
to be given by $\hat{G}_{\mu\nu}$ as 
\begin{equation}
\hat{F}_{\mu\nu}=\frac{1}{2ir^{2}}\hat{G}_{\mu\nu}, 
\label{matrixversionofF}
\end{equation}
where $r^{2}=\alpha^{2}n(n+2k)$. 
The field strength on a (commutative) $2k$-sphere 
is originally given in (\ref{fieldstrengthspin(2k+1)}). 
The equation (\ref{matrixversionofF}) manifests that 
the field strength is completely expressed by a matrix. 
In other words, a noncommutative description of 
(\ref{fieldstrengthspin(2k+1)}) is (\ref{matrixversionofF}). 
The coordinate dependence has disappeared, 
and the information of the coordinate is given by 
the eigenvalue of the matrix. 
The index of the matrix represents 
not only the color space but also the coordinate on the sphere. 
The noncommutativity of the coordinate and that of 
the nonabelian gauge field are mixed.  
We can no longer distinguish them. 
Noncommutative geometry mixes the two kinds of space 
in an interesting way. 
These facts can be summarized in the following sentences. 
A fuzzy $2k$-sphere is provided by an $N_{k}\times N_{k}$ 
matrix. This means that 
$N_{k}$ quanta (or D-particles) form a fuzzy $2k$-sphere. 
The space of a nonabelian gauge field is formed by 
$N_{k-1}$ quanta, and $N_{k}/N_{k-1}$ represents the number 
of points on the sphere. The matrix space of 
a fuzzy $2k$-sphere has a structure such as locally 
$S^{2k}\times$ fuzzy $S^{2k-2}$, where 
the fuzzy $S^{2k-2}$ construct a nonabelian gauge field. 
This structure was suggested 
in \cite{hep-th/0111278,hep-th/0301055} 
from the algebraic point of view.

\vspace{0.4cm}

In the latter part of this section, we consider 
a large $n$ limit 
which relates a fuzzy sphere with a fuzzy plane. 
Since the radius $r$ is related to $n$, 
we can make $r$ large by taking a large $n$. 
In this limit, a fuzzy sphere looks like a fuzzy plane 
and we can formally obtain a fuzzy plane. 
Such a large $n$ limit 
for the fuzzy sphere algebra was discussed in \cite{hep-th/0301055}. 
The fuzzy sphere algebra basically reduces to 
some sets of the Heisenberg algebra 
($\sim$ two-dimensional noncommutative plane). 
Since the fuzzy $2k$-sphere is a $k(k+1)$-dimensional 
space, $k(k+1)/2$ sets of the Heisenberg algebra 
are obtained. 
We now confirm that the field strength reduces to 
abelian (or commutative)  
because a two-dimensional noncommutative plane 
is realized by introducing an abelian gauge field. 
Since the field strength is given in terms of 
coordinates of (lower) dimensional fuzzy sphere 
as in (\ref{fieldstrengthgeneral}), 
we can use the calculation of \cite{hep-th/0301055}. 
We show only results. 
For the $k=2$ case, the field strength reduces to 
\begin{equation}
F_{12}=-F_{34}\simeq \frac{n}{2r^{2}}1
=\frac{1}{2l_{nc}^{2}}1,
\end{equation}
and other components are zero. 
We thus obtained two $U(1)$ gauge fields. 
Nonzero components of the field strength for 
the $k=3$ and $k=4$ cases can also be shown as 
\begin{equation}
F_{12}=-F_{34}=-F_{56}\simeq \frac{n}{2r^{2}}1
=\frac{1}{2l_{nc}^{2}}1
\end{equation}
and 
\begin{equation}
F_{12}=-F_{34}=-F_{56}=-F_{87}\simeq \frac{n}{2r^{2}}1
=\frac{1}{2l_{nc}^{2}}1
\end{equation}
respectively. As expected from the correspondence 
between a noncommutative plane and an abelian gauge field, 
the nonabelian gauge fields have reduced to some abelian ones. 
The charges of lower dimensional branes 
no longer vanish after we take this large $n$ limit. 


\section{Summary and Discussions}
\label{sec:summarydiscussions}
\hspace{0.4cm}
To describe higher dimensional D-branes 
by using lower dimensional D-branes, 
we need noncommutative geometry. 
This description is closely related to 
the matrix model description of D-branes. 
It is known that such a higher dimensional D-brane 
is not just pure a D-brane but 
a bound state of a higher dimensional D-brane 
and lower dimensional D-branes. 
Lower dimensional D-branes bounded on them are expressed by 
a gauge field configuration 
with nonvanishing Chern characters. 
In this sense, gauge field configurations on 
D-branes are dual to noncommutative geometry. 
To understand the dual description is directly connected with 
to understand noncommutative geometry. 
We expect that such studies help us 
to know the role of noncommutative geometry 
in string theory. 

To consider such relationships not only for 
flat D-branes but also for curved ones is an important subject. 
A fuzzy sphere is used in matrix models to construct 
a spherical geometry. 
We can interpret it as a bound state of spherical 
D-branes and D$0$-branes. 
D$0$-branes on higher dimensional spherical D-branes 
are regarded by a nonabelian gauge configuration. 
Higher dimensional fuzzy spheres have some unusual features. 
One of them is that the number of the dimension of 
fuzzy spheres is different from 
that of usual spheres. The dimension of a fuzzy $2k$-sphere 
is $k(k+1)$. $k(k-1)$ dimensions are clearly extra compared 
to a usual $2k$-sphere. 
The role of them for constructing noncommutative geometry 
is essential. 
The origin of them has been considered as the use of 
nonabelian gauge fields. 
In this paper, we considered two description for 
a bound state of D$0$-branes and D$2k$-branes 
in a constant R-R $(2k+2)$-form field strength background. 
The first is the D$0$-brane description, and  
the second is the D$2k$-brane description. 
A fuzzy $2k$-sphere appears as a classical solution of 
a matrix model of D$0$-branes. 
A dual description of this is 
obtained by introducing nonabelian gauge fields.  
We compared some quantities such as the values of the potential, 
the radius of sphere and lower dimensional brane charges 
for these two descriptions. 
These two descriptions provide different results 
because each description can be trusted in 
different parameter regions. 
Taking a large $n$ limit leads to the agreement 
of various quantities including the coefficients. 
We provided an explanation for the large $n$ limit 
by considering the analogy to a quantum Hall system. 
The large $n$ limit can be interpreted as 
the lowest Landau level condition. 
We finally arrived at the conclusion that 
a fuzzy $2k$-sphere 
is the same object as $N_{k-1}$ D$2k$-branes 
with an $SO(2k)$ gauge field in the limit of large $n$. 
When $n$ is large, we can relate commutative variables 
with noncommutative variables. 
Not only the coordinates 
on spherical D-branes but also a nonabelian gauge field strength 
are expressed by noncommuting matrices. 

The fuzzy sphere algebra is composed of two kinds of matrix, 
$\hat{G}_{\mu}$ and $\hat{G}_{\mu\nu}$. 
Although both form the noncommutative algebra 
(\ref{so(2k+1)algebra}) and (\ref{fuzzyspherealgebra2}) 
and are treated on the same footing, 
the origin of noncommutativity of them is different.  
Noncommutativity of $\hat{G}_{\mu}$ is due to the 
existence of a magnetic field. 
On the other hand, that of $\hat{G}_{\mu\nu}$ 
comes from the nonabelian property 
of the field strength. 

Investigating the dynamics of curved D-branes is an important subject. 
The fuzzy two-sphere shows an interesting classical dynamics. 
Reducible representations of $SU(2)$ 
including separated D$0$-branes 
are unstable, condensing to an irreducible representation 
as a stable state \cite{hep-th/0110172}.
\footnote{This phenomenon is investigated from 
the viewpoint of the tachyon condensation in 
\cite{hep-th/0309082}.} 
It is, however, not expected to see a similar phenomenon 
in the case of higher dimensional fuzzy spheres. 
A big difference which distinguishes higher dimensional 
fuzzy sphere from fuzzy two-sphere 
is that 
$N_{1}$ can take any integers while $N_{k}(k\neq1)$ cannot. 
The fact that the size of the matrix is limited  
is due to the use of higher dimensional algebra
\footnote{
This is also true for a noncommutative $CP^{2}$, 
which is constructed from the $SU(3)$ algebra.}.  
This is related to the fact that the coefficient 
of the coupling to the R-R field strength 
depends on $n$ (see (\ref{explicitvalueoff})). 
These facts which cannot be seen in the case of 
fuzzy two-sphere would restrict the dynamics of higher dimensional 
noncommutative branes. 
Another difference is found by seeing the equation 
(\ref{D0potentialvalue}), which shows 
that classical energy of higher dimensional spheres is
higher than that of D$0$-branes. 
Accordingly higher dimensional fuzzy spheres have 
a classical instability. 
Taking account of these facts, 
to study the dynamics of higher dimensional 
fuzzy spheres including quantum corrections 
\cite{hep-th/0303120,hep-th/0401038,hep-th/0401120} 
is an interesting future subject. 

It is also interesting to investigate a dual description 
for another noncommutative curved brane. 
It would be worth while examining a noncommutative $CP^{2}$, 
which is realized by the $SU(3)$ algebra. 
Some extra dimensions also exist in noncommutative $CP^{2}$, 
which is studied from the algebraic point of view 
in \cite{hep-th/0309082}. Exactly speaking, 
the existence of them depends on the representation of $SU(3)$. 
The choice of the representation 
is related to the choice of the gauge group 
in a dual description. 
The construction of the quantum Hall system 
on $CP^{2}$ \cite{hep-th/0203264} 
suggests this relationship.

\vspace{1.0cm}
\begin{center}
{\bf Acknowledgments}
\end{center}
\hspace{0.4cm}

I would like to express my gratitude to K. Hasebe 
for helpful discussions.

\renewcommand{\theequation}{\Alph{section}.\arabic{equation}}
\appendix

\section{Some Formulae of Fuzzy Sphere}
\label{sec:someformulaefuzzysphere}
\setcounter{equation}{0} 
\hspace{0.4cm}
In this appendix, we summarize several formulae 
involving 
$\hat{G}_{\mu}$ and $\hat{G}_{\mu\nu}$ 
in diverse dimensions. 
The dimension $N_{k}$ is given by 
\begin{eqnarray}
&&N_{1}=n+1, \hspace{0.4cm}
N_{2}=\frac{1}{6}(n+1)(n+2)(n+3), \cr 
&&N_{3}=\frac{1}{360}(n+1)(n+2)(n+3)^{2}(n+4)(n+5), \cr
&&N_{4}=\frac{1}{302400}(n+1)(n+2)(n+3)^{2}
(n+4)^{2}(n+5)^{2}(n+6)(n+7), 
\end{eqnarray}
where $n$ is a positive integer. 
We have the following relations 
\begin{equation}
\hat{G}_{\mu}\hat{G}_{\mu}=n(n+2k)  
\label{GmuGmu=nn}
\end{equation}
and 
\begin{equation}
\hat{G}_{\mu\nu}\hat{G}_{\nu\mu}=2kn(n+2k). 
\label{GmunuGmunu=nn}
\end{equation} 
The following relations are also satisfied 
\begin{equation}
\hat{G}_{\mu\nu}\hat{G}_{\nu}
=2k\hat{G}_{\mu}
\label{gmunugnu=gmu}
\end{equation}
and 
\begin{equation}
\hat{G}_{\mu\nu}\hat{G}_{\nu\lambda} 
=n(n+2k)\delta_{\mu\lambda}
 +(k-1)\hat{G}_{\mu}\hat{G}_{\lambda}
 -k\hat{G}_{\lambda}\hat{G}_{\mu}.
\end{equation}
$G_{\mu}$ satisfy the following relation 
\begin{equation}
\epsilon^{\mu_{1}\cdots\mu_{2k}\mu_{2k+1}}
\hat{G}_{\mu_{1}}
\cdots\hat{G}_{\mu_{2k}}
=C_{k}\hat{G}_{\mu_{2k+1}}
\label{appnoncommurelationG}
\end{equation}
where $\epsilon^{\mu_{1}\cdots\mu_{2k}\mu_{2k+1}}$ 
is the $SO(2k+1)$ invariant tensor. 
$C_{k}$ is a constant which depends on $n$,   
\begin{equation}
C_{1}=2i, \hspace{0.2cm}
C_{2}=8(n+2), \hspace{0.2cm}
C_{3}=-48i(n+2)(n+4), \hspace{0.2cm}C_{4}=-384(n+2)(n+4)(n+6).
\label{valueofC}
\end{equation}
The details of this calculation are given in 
\cite{azumabagnoud}. 
By multiplying the equation (\ref{appnoncommurelationG}) 
by $\hat{G}_{\mu_{2k+1}}$, 
we have 
\begin{eqnarray}
\epsilon^{\mu_{1}\cdots \mu_{2k}\mu_{2k+1}}
\hat{G}_{\mu_{1}} \cdots 
\hat{G}_{\mu_{2k}}\hat{G}_{\mu_{2k+1}}
&=&C_{k}n(n+2k) \cr  
&=&\frac{i}{2(k+1)}nC_{k+1}, 
\label{noncommurelationG2}
\end{eqnarray}
where we have used the following relation 
which is found from (\ref{valueofC}), 
\begin{equation}
C_{k}
=-i2k(n+2k-2)C_{k-1}. 
\label{valuesofCk}  
\end{equation}


\section{Hopf map and Berry phase}
\label{sec:hopfmapandberryphase}
\hspace{0.4cm}
In this section, we review the construction of 
the monopole gauge field on an even-dimensional sphere 
\cite{CNYang,horvathpalla} 
by using the Hopf map 
\cite{cond-mat/9805404,zhanghu,cond-mat/0306045,hep-th/0310274}.

\vspace{0.4cm}

We now consider the first Hopf map. 
It is known as a map from $S^{3}$ to $S^{2}$, and 
naturally introduces a $U(1)$ bundle on $S^{2}$. 
We prepare the following projection operator, 
\begin{equation}
P=\frac{1}{2}(1_{2}+n_{\mu}\sigma_{\mu})=
 \left( \begin{array}{c c}
  1+n_{3} & n_{1}-in_{2} \\ 
 n_{1}+in_{2}  & 1-n_{3}   \\
 \end{array} \right),  
\end{equation}
which satisfies $P^{2}=P$. 
$x_{\mu}=rn_{\mu}$ is a coordinate of $S^{2}$. 
The eigenstate of $P$ is given by 
\begin{equation}
|v_{N}\rangle =\frac{1}{N_{N}}P
 \left( \begin{array}{c}
  1 \\ 0   \\
 \end{array} \right) = 
 \frac{1}{N_{1}}
 \left( \begin{array}{c}
  r+x_{3}  \\ 
 x_{1}+ix_{2}    \\
 \end{array} \right), 
\end{equation}
where $N_{N}=\sqrt{2r(r+x_{3})}$ is a normalization factor, which 
ensures $\langle v|v \rangle =1$.  
The Berry phase is defined \cite{cond-mat/9805404} as 
\begin{eqnarray}
\gamma_{N}&=&-i\int_{0}^{t}d\tau\langle v(\tau)|\frac{d}{d\tau}
|v (\tau)\rangle 
=-i\int_{0}^{t}\langle v|d|v\rangle  \cr 
&=& \int_{0}^{t} A_{N}^{\mu}dx_{\mu}. 
\end{eqnarray}
$A_{\mu}^{N}$ is the Dirac monopole: 
\begin{equation}
A_{N}=A_{N}^{\mu}dx_{\mu}=\frac{1}{2r(r+x_{3})}
(x_{1}dx_{2}-x_{2}dx_{1})=
\frac{1}{2r(r+x_{3})}
\epsilon_{ab}x_{a}dx_{b}, 
\end{equation}
which is singular at $x_{3}=-r$. 
A monopole solution which is singular at $x_{3}=r$ 
is obtained by replacing $|v_{N}\rangle$ with 
\begin{equation}
|v_{S}\rangle =\frac{1}{N_{S}}P
 \left( \begin{array}{c}
  0 \\ 1   \\
 \end{array} \right),
\end{equation}
where $N_{S}=\sqrt{2r(r-x_{3})}$. 
The field strength of the Dirac monopole is calculated as 
\begin{equation}
F_{\mu\nu}=\partial_{\mu}A_{\nu}-\partial_{\nu}A_{\mu}=
\frac{1}{2r^{3}}\epsilon_{\mu\nu\rho}x^{\rho}. 
\end{equation}

%

\vspace{0.4cm}

We next consider the second Hopf map: $S^{7}\rightarrow S^{4}$. 
This gives an $SU(2)$ bundle on $S^{4}$. 
The construction can be done in the same way as in the first Hopf map. 
The projection operator we need in this case is 
\begin{equation}
P=\frac{1}{2}(1_{4}+n_{\mu}\Gamma_{\mu})=
 \left( \begin{array}{c c}
  1+n_{5} & n_{4}-in_{i}\sigma_{i} \\ 
 n_{4}+in_{i}\sigma_{i}  & 1-n_{5}   \\
 \end{array} \right). 
 \label{projectionS^{4}} 
\end{equation}
Our notation of the five-dimensional 
gamma matrix is  
\begin{eqnarray} 
\Gamma_{i}  
 &=&   \left( \begin{array}{c c}
  0 & -i\sigma_{i} \\ 
 i\sigma_{i}  & 0   \\
 \end{array} \right), \hspace{0.2cm}(i=1,2,3)     \cr
\Gamma_{4}  
 &=&   \left( \begin{array}{c c}
  0 & 1_{2} \\ 
  1_{2}  & 0   \\
 \end{array} \right) ,     \hspace{0.4cm}
\Gamma_{5}  
 =   \left( \begin{array}{c c}
  1_{2} & 0 \\ 
 0  & -1_{2}   \\
 \end{array} \right),   
 \label{fivedimensionalgammamatrix}     
\end{eqnarray}
where $\sigma_{i}$ is the Pauli matrices. 
They satisfy the Clifford algebra 
\begin{equation}
\{\Gamma_{\mu} ,\Gamma_{\nu} \}=2\delta_{\mu\nu}.
\end{equation}
The $SU(2)$ gauge potential is obtained 
by calculating the Berry phase, 
\begin{eqnarray}
A_{N}&=&A_{N}^{\mu}dx_{\mu}=\frac{1}{2r(r+x_{5})}
(\sigma_{ij}x_{i}dx_{j}
+\sigma_{i}x_{4}dx_{i}-\sigma_{i}x_{i}dx_{4}) \cr 
&=&-\frac{1}{2r(r+x_{5})}\eta_{\alpha\beta}^{i}
x_{\beta}\sigma_{i}dx_{\alpha}, 
\label{SU(2)gaugefieldnorthpole}
\end{eqnarray}
where 
$\sigma_{ij}\equiv\frac{1}{2i}[\sigma_{i},\sigma_{j}]
=\epsilon_{ijk}\sigma_{k}$. 
This is called the Yang monopole \cite{CNYang}. 
This is singular at the south pole $x_{5}=-r$. 
We have introduced the t'Hooft symbol; 
\begin{eqnarray}
\eta_{\alpha\beta}^{i} = \epsilon_{i\alpha\beta4} 
- \delta_{i\alpha}\delta_{4\beta} 
+ \delta_{i\beta}\delta_{4\alpha},  
\end{eqnarray}
where the nonzero values are  
$\eta_{23}^{1}=\eta_{41}^{1}=\eta_{31}^{2}=
\eta_{42}^{2}=\eta_{12}^{3}=\eta_{43}^{3}=1$.
We note that 
$\Sigma_{\mu\nu}^{N}\equiv i\eta_{\mu\nu}^{i}\sigma_{i}$ satisfy 
\begin{equation}
[\Sigma_{\mu\nu}^{N},\Sigma_{\lambda\rho}^{N}]=2\left(
\delta_{\nu\lambda}\Sigma_{\mu\rho}^{N}
+\delta_{\mu\rho}\Sigma_{\nu\lambda}^{N}
-\delta_{\mu\lambda}\Sigma_{\nu\rho}^{N}
-\delta_{\nu\rho}\Sigma_{\mu\lambda}^{N}
\right). 
\label{rotationalgebra}
\end{equation}
$\Sigma_{\mu\nu}^{N}$ is actually a left upper part of 
the matrix $\Gamma_{\mu\nu}$ $(\mu,\nu=1,\cdots,4)$, 
where $\Gamma_{\mu}$ is given by (\ref{fivedimensionalgammamatrix}).  
We define the field strength as follows, 
\begin{eqnarray}
F_{\mu\nu}&\equiv& -i[D_{\mu},D_{\nu}]
\equiv -i[\partial_{\mu}+iA_{\mu},\partial_{\nu}+iA_{\nu}] \cr
&=&\partial_{\mu}A_{\nu}-\partial_{\nu}A_{\mu}
+i[A_{\mu},A_{\nu}]. 
\end{eqnarray}
The field strength for the Yang monopole 
(\ref{SU(2)gaugefieldnorthpole}) is 
\begin{eqnarray}
&&F^{N}_{5\alpha}=-\frac{r+x_{5}}{r^{2}}A_{\alpha} \cr
&&F^{N}_{\alpha\beta}=\frac{1}{r^{2}}\left(A_{\alpha}x_{\beta}
-A_{\beta}x_{\alpha}
+\frac{1}{2}\eta_{\alpha\beta}^{i}\sigma_{i}
\right), 
\label{SU(2)fieldstrength}
\end{eqnarray}
where $\alpha,\beta=1,\cdots,4$.

\vspace{0.4cm}


We next consider the monopoles on $S^{6}$ and $S^{8}$. 
As a generalization of the monopoles on $S^{2}$ and $S^{4}$, 
it is natural to start with the 
following projection operator 
\begin{equation}
P^{(k)}\equiv\frac{1}{2}(1_{2^{k}}+n_{\mu}\Gamma_{\mu})=
 \left( \begin{array}{c c}
  1+n_{2k+1} & n_{2k}-in_{i}\gamma_{i} \\ 
 n_{2k}+in_{i}\gamma_{i}  & 1-n_{2k+1}   \\
 \end{array} \right),   
\end{equation}
where $k=3,4$. 
An explicit form of the 
$2^{k}\times 2^{k}$ $(2k+1)$-dimensional gamma matrix is 
\begin{eqnarray} 
\Gamma_{i}  
 &=&   \left( \begin{array}{c c}
  0 & -i\gamma_{i} \\ 
 i\gamma_{i}  & 0   \\
 \end{array} \right), \hspace{0.2cm}(i=1,\ldots,2k-1)     \cr
\Gamma_{2k}  
 &=&   \left( \begin{array}{c c}
  0 & 1_{2^{k-1}} \\ 
  1_{2^{k-1}}  & 0   \\
 \end{array} \right) ,     \hspace{0.4cm}
\Gamma_{2k+1}  
 =   \left( \begin{array}{c c}
  1_{2^{k-1}} & 0 \\ 
 0  & -1_{2^{k-1}}   \\
 \end{array} \right),  
 \label{2k+1dimgammamatrix}      
\end{eqnarray}
where $\gamma_{i}$ is the $(2k-1)$-dimensional gamma matrix. 
This map was used in \cite{hep-th/0310274} to 
construct a higher dimensional quantum Hall system.  
One may notice that the map for $k=4$ case is different 
from the third Hopf map. 
The calculation of the Berry phase 
introduces the following $SO(2k)$ monopole field on 
$S^{2k}$: 
\begin{eqnarray}
A_{N}=A_{N}^{\mu}dx_{\mu}
&=&-i\langle v|d|v\rangle
=-i \left( \begin{array}{c c}
  1  & 0\\ 
 \end{array} \right) \frac{P^{(k)}}{N^{(k)}}
d\left(\frac{P^{(k)}}{N^{(k)}}\right)
 \left( \begin{array}{c}
  1 \\ 0   \\
 \end{array} \right) \cr
&=&\frac{1}{2r(r+x_{2k+1})}
\left(-i\gamma_{ij}x_{i}dx_{j}
+\gamma_{i}x_{2k}dx_{i}-\gamma_{i}x_{i}dx_{2k}\right)  \cr
&=&\frac{-i}{2r(r+x_{2k+1})}
\Sigma_{\alpha\beta}^{N}x_{\alpha}dx_{\beta}
\label{spin2kgaugefield}
\end{eqnarray}
where 
$\Sigma_{\alpha\beta}^{N}=(\Sigma_{ij}^{N},\Sigma_{2k,i}^{N})
\equiv (\gamma_{ij},i\gamma_{i})
=(\frac{1}{2}[\gamma_{i},\gamma_{j}],i\gamma_{i})$ 
is the spinor representation of $SO(6)$, 
satisfying the algebra of (\ref{rotationalgebra}). 
$\Sigma_{\alpha\beta}^{N}$ 
is a subspace of $\Gamma_{\alpha\beta}$ $(\alpha,\beta=1,\cdots,2k)$ 
which is labelled by $\Gamma_{2k+1}=1$, 
where $\Gamma_{\mu}$ is the $(2k+1)$-dimensional gamma 
matrix (\ref{2k+1dimgammamatrix}).  
%
%
The gauge field strength corresponding to the above 
monopole gauge field is 
\begin{eqnarray}
&&F^{N}_{2k+1\alpha}=-\frac{r+x_{2k+1}}{r^{2}}A_{\alpha} \cr
&&F^{N}_{\alpha\beta}=\frac{1}{r^{2}}
\left(A_{\alpha}x_{\beta}-A_{\beta}x_{\alpha}
-\frac{i}{2}\Sigma_{\alpha\beta}^{N}
\right),  
\label{fieldstrengthspin(2k+1)}
\end{eqnarray}
where $\alpha,\beta=1,\cdots,2k$. 
These monopole fields have the $SO(2k+1)$ symmetry. 

\vspace{0.4cm}

At the north pole ($x_{\alpha}=0$, $x_{2k+1}=r$), 
the field strength becomes 
$F_{2k+1\alpha}=0$, $F_{\alpha\beta}
=-\frac{i}{2r^{2}}\Sigma_{\alpha\beta}^{N}$. 
The field strength on $S^{2k}$ $(k=2,3,4)$ at the north pole 
satisfies the following relations, 
\begin{eqnarray}
&& \epsilon_{\alpha\beta\gamma\delta}F^{\gamma\delta}
=-2F_{\alpha\beta}, \cr 
&& \epsilon_{\alpha_{1}\alpha_{2}\alpha_{3}\alpha_{4}\alpha_{5}\alpha_{6}}
F^{\alpha_{3}\alpha_{4}}F^{\alpha_{5}\alpha_{6}}=
\frac{4}{r^{2}}(n+2)F_{\alpha_{1}\alpha_{2}}, \cr 
&& \epsilon_{\alpha_{1}\alpha_{2}\alpha_{3}\alpha_{4}
\alpha_{5}\alpha_{6}\alpha_{7}\alpha_{8}}
F^{\alpha_{3}\alpha_{4}}F^{\alpha_{5}\alpha_{6}}
F^{\alpha_{7}\alpha_{8}}=
-\frac{12}{r^{4}}(n+2)(n+4)F_{\alpha_{1}\alpha_{2}} .
\end{eqnarray}
The first one is a (anti)self-dual relation of the instanton, 
and the second and the third ones are considered as 
a higher dimensional generalization of the self-dual relation.


\end{document}